\documentclass[12pt,preprint]{aastex}
\usepackage{natbib}
\newcommand{\msini}{\ensuremath{m \sin{i}}}
\newcommand{\feh}{\ensuremath{[\mbox{Fe}/\mbox{H}]}}

\newcommand{\mv}{\ensuremath{M_{\mbox{\scriptsize V}}}}

\newcommand{\dmv}{\ensuremath{\Delta\mv}}
\newcommand{\teff}{\ensuremath{T_{\mbox{\scriptsize eff}}}}
\newcommand{\persec}{\ensuremath{\mbox{s}^{-1}}}
\newcommand{\mps}{\mbox{m} \persec}
\newcommand{\mjup}{\ensuremath{\mbox{M}_{\mbox{Jup}}}}

\newcommand{\Mjupsmall}{\mbox{\scriptsize M}_{\mbox{\tiny Jup}}}
\def\astrosun {\mbox{$\odot$}}
\newcommand{\Msol}{\ensuremath{\mbox{M}_{\astrosun}}}

\shorttitle{Long Period \& Low Amplitude Exoplanet Candidates}
\shortauthors{Wright et al.}

\begin{document}
\title{Four New Exoplanets, and Hints of Additional Substellar Companions to Exoplanet Host Stars\altaffilmark{1}}
\altaffiltext{1}{Based on observations obtained
at the W. M. Keck Observatory, which is operated jointly by the
University of California and the California Institute of Technology.
The Keck Observatory was made possible by the generous financial
support of the W. M. Keck Foundation.}
\author{J. T. Wright\altaffilmark{2}, G. W. Marcy\altaffilmark{2}, 
 D. A Fischer\altaffilmark{3}, R. P. Butler\altaffilmark{4}, 
 S. S. Vogt\altaffilmark{5}, C. G. Tinney\altaffilmark{6}, 
 H. R. A. Jones\altaffilmark{7}, B. D. Carter\altaffilmark{8}, 
 J. A. Johnson\altaffilmark{2}, C. McCarthy\altaffilmark{3}, 
 K. Apps\altaffilmark{9}}

\altaffiltext{2}{Department of Astronomy, 601 Campbell Hall, University of California, Berkeley, CA 94720-3411}
\altaffiltext{3}{Department of Physics and Astronomy, San Francisco State University, San Francisco, CA 94132}
\altaffiltext{4}{Department of Terrestrial Magnetism, Carnegie Institute of Washington, 5241 Broad Branch Road NW, Washington, DC 20015-1305}
\altaffiltext{5}{UCO/Lick Observatory, University of California, Santa Cruz, CA 95064}
\altaffiltext{6}{Anglo-Australian Observatory, PO Box 296, Epping. 1710. Australia}
\altaffiltext{7}{Centre for Astrophysics Research, University of
  Hertfordshire, Hatfield, AL 10 9AB, England, UK}
\altaffiltext{8}{Faculty of Sciences, University of Southern Queensland, Toowoomba. 4350. Australia}
\altaffiltext{9}{Physics and Astronomy, University of Sussex, Falmer,
  Brighton, BN1 9QJ, England, UK}

\begin{abstract}

We present four new exoplanets:  HIP 14810 b \& c, HD 154345 b, and HD
187123 c.  The two planets orbiting HIP 14810, from the N2K project,
have masses of 3.9 and 0.76 \mjup.  We have searched the radial
velocity time series of 90 known exoplanet 
systems and found new residual trends due to additional, long
period companions.  Two stars known to host one exoplanet have sufficient
curvature in the residuals to a one planet fit to constrain the
minimum mass of the outer companion to be substellar: HD 68988 c with
$8\,\mjup < \msini < 20 \,\mjup$ and HD 187123 c with $3 \,\mjup <
\msini < 7 \,\mjup$, both with $P > 8$ y.  We have also searched the
velocity residuals of known exoplanet systems for prospective
low-amplitude exoplanets and present some candidates.   We discuss
techniques for constraining the mass and period of exoplanets in such
cases, and for quantifying the significance of weak RV signals. We also present
two substellar companions with incomplete orbits and periods longer
than 8 y:  HD 24040 b and HD 154345 b with $\msini < 20 \mjup$ and
$\msini < 10$ \mjup, respectively.  

\end{abstract}


\section{Introduction}
Of the 151 nearby stars known to harbor one or more planets, 19 are
well-characterized multiple-planet systems, and an additional 24 show
radial
 velocity (RV) residuals 
indicative of additional companions \citep{Butler06}.  For instance,
\citet{Vogt05} reported additional companions around five stars,
including two revealed by incomplete orbits apparent in the
RV residuals (HD 50499 and HD 217107), and one as a
short-period, low amplitude variation in the residuals of the fit to a
long-period outer companion.  \citet{Rivera05} detected a 7.5 Earth-mass
mass companion to GJ 876 in a 2-day period through analysis of the RV
residuals to a 2-planet dynamical fit of the more massive, outer exoplanets.
\citet{Gozdziewski06} similarly analyzed the RV residuals of 4 stars
to search for Neptune-mass companions.  

Very little is known about the frequency or nature of exoplanets
with orbital distances greater than 5 AU \citep{Marcy05}.
Precise radial velocities have only reached the precision required
to detect such objects within the last 10 years \citep{Butler96b},
which is less than the orbital period of such objects ($P>12$ y for
exoplanets orbiting solar mass stars).  Thus, the RV curves for such
planets are all necessarily incomplete, and we must obtain many more
years of data before our knowledge of their orbits improves
significantly.  

The ability to put constraints on planets with incomplete
orbits, however weak, allows us to peek beyond the 5 AU completeness
limit inherent in the ten-year-old planet searches.  Characterizing
incomplete orbits also increases our sample of known 
multiple exoplanetary systems, which improves our understanding of the
frequency of orbital resonances, the growth of multiple planets, and
the mechanics of orbital migration.   

In this work, we present our analysis of the RV data of \citet{Butler06} in an effort
to determine which of those systems have additional, low-amplitude
companions.

Many systems known to host one exoplanet show more distant, long-period companions with highly
significant but incomplete orbits.  In these systems, it can be
extremely difficult to constrain the properties of the outer
companion:  in the case of a simple trend with no curvature, very
little can be said about the nature of these companions beyond their
existence, but even this informs studies of exoplanet multiplicity and
the frequency of exoplanets in binary systems.

In \S~\ref{hip14810} we discuss a new multiple planet system
from the N2K project, HIP 14810.  In
\S~\ref{trends} we describe how we have employed
a false alarm probability statistic to test the significance of trends in the RV data of
stars already known to host exoplanets.  We find that six stars known
to host exoplanets have previously undetected trends, and thus
additional companions. 

When the RV residuals to a single Keplerian show significant
curvature, one may be able to place additional constraints on the
maximum $\msini$ of the additional companion.  In \S\S~\ref{mapsection}--\ref{twonew} we present our analysis
of this problem in the cases of HD 24040 b and HD 154345 b, two
substellar companions new to this work with very incomplete orbits.  By mapping
$\chi^2$ space for Keplerian fits, we show that HD 154345 b is
almost certainly planetary ($\msini <10 \mjup$), and that HD 24040 b may
be planetary ($5\mjup < \msini < 30\mjup$).  

In \S~\ref{mining} we describe how we extended this method to the RV residuals of known
planet-bearing stars which show trends.  We find that for 2 stars we can place sufficiently strong upper limits on \msini\
to suggest that the additional companions are planetary in nature.

\section{HIP 14810}
\label{hip14810}
\begin{figure} 
  \plotone{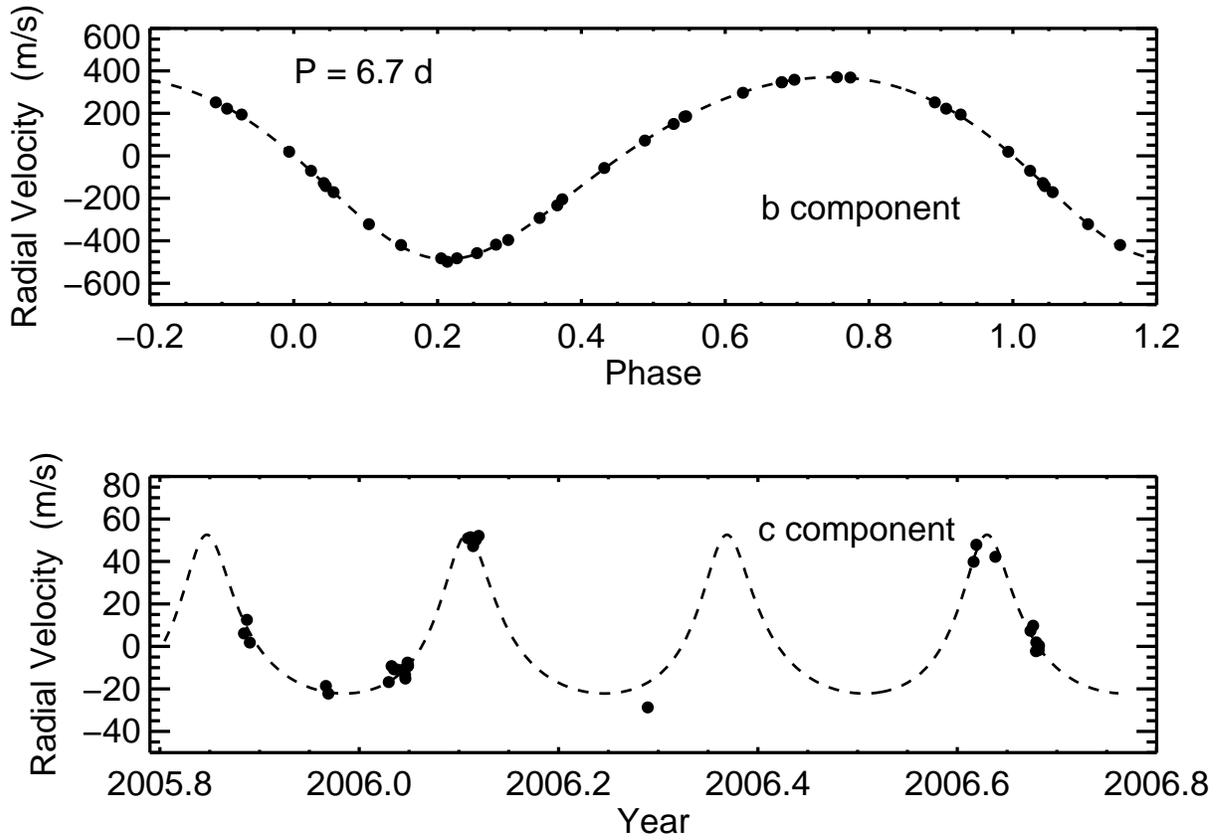}
  \caption{RV curve for HIP 14810 with data from Keck, showing the inner planet with $P =
  6.67$ d and $\msini=3.9 \mjup$ the outer planet with $P=95.3$ d and
  $\msini =0.76$ 
  \mjup.\label{hip14810fig}} 
\end{figure}

HIP 14810 is a metal-rich ($\feh = 0.23$) G5 V, V=8.5 star which we have observed at Keck
Observatory as part of the N2K program \citep{Fischer05} since Nov
2005.  Table~\ref{RVhip14810} contains the RV data for this star.  Its
stellar characteristics are listed in Table~\ref{starchar}, 
determined using the same LTE spectral analysis used for stars in the
SPOCS catalog \citep{Valenti05}.  We quickly detected a 
short-period, high-amplitude companion ($P=6.67$ d, $\msini=3.9
\,\mjup$) and a strong, $\sim 200$ m/s trend.  Further observations
revealed evidence for substantial curvature in the residuals to a
planet plus trend fit.  Fig.\ref{hip14810fig} shows the RV curve for this
star decomposed into Keplerian curves for the b and c components, and
Table~\ref{orbitupdates} contains the best-fit double Keplerian elements.

\begin{deluxetable}{lrc}
\tablecaption{RV Data for HIP 14810\label{RVhip14810}}
\tablecolumns{3}
\tablewidth{0pc}
\tablehead{
{Time} &{Radial Velocity} & {Unc.}\\ 
{(JD-2440000)} &{(m/s)}&{(m/s)}}

\startdata
13693.760579 & -130.8 &1.3  \\
13694.831481 & -473.6 &1.2    \\
13695.909225 & -226.9 &1.2    \\
13723.786250 & 162.6 &1.0     \\
13724.688484 & 324.9 &1.2     \\
13746.814595 & 2.4 &1.3       \\
13747.852940 & -435.82 &0.94  \\
13748.734190 & -433.3 &1.2    \\
13749.739236 & -71.4 &1.2     \\
13751.898252 & 358.3 &1.1     \\
13752.807431 & 241.05 &0.80   \\
13752.912477 & 211.6 &1.7     \\
13753.691574 & -79.8 &1.1     \\
13753.810359 & -137.6 &1.2    \\
13753.901042 & -180.2 &1.2    \\
13775.836157 & -240.01 &0.97  \\
13776.812859 & 123.4 &1.4     \\
13777.723102 & 346.7 &1.3     \\
13778.720799 & 416.0 &1.3     \\
13779.744410 & 238.4 &1.3     \\
13841.722049 & -515.7 &1.4    \\
13961.130301 & -280.9 &1.0    \\
13962.133333 & -413.4 &1.1    \\
13969.097315 & -348.3 &1.2    \\
13981.969815 & -476.5 &1.2    \\
13982.947431 & -200.9 &1.2    \\
13983.981470 & 151.5 &1.0     \\
13984.096979 & 187.3 &1.2     \\
13984.985775 & 345.7 &1.3     \\
13985.102106 & 357.6 &1.3     
\enddata

\end{deluxetable}

A 2-planet Keplerian fit yields an outer planet with $\msini = 0.95\,\mjup$,
$P=114$ d, and eccentricity of 0.27.  We present the orbital solutions for 
this two-planet system in Table~\ref{orbitupdates}.  

\section{Detecting Long-Period Companions}
\label{trends}
Very long period substellar companions appear in radial
velocity data first as linear trends (constant accelerations), then
as trends with curvature, and finally, as the duration of the
observations becomes a substantial fraction of the orbital period, as
recognizable portions of a Keplerian velocity curve.  It is important,
then, to have a statistically robust test for trends in
velocity residuals.  In this section, we discuss calculating false alarm
probabilities (FAPs) for such trends.

\subsection{Using FAP to Detect Trends\label{FAPtrend}}
\citet[\S~5.2][]{Marcy05} present a detailed discussion of using false
alarm probabilities (FAPs) for determining the
significance of a periodic signal in an RV time series.  Here, our task
is similar.  We wish to test the hypothesis that a star has an
additional companion with a long period, manifest only as a linear
trend in the RV series.  We compare this hypothesis to the null
hypothesis that the data are adequately described only by the best-fit
Keplerians and noise.  

We first fit the data set with a Keplerian model and compare the
$\chi^2_{\nu}$ statistic to that of a model employing a Keplerian plus a
linear trend.  If this statistic improves, that is, if
$\Delta\chi^2_{\nu} = \chi^2_{\nu,\mbox{trend}}-\chi^2_{\nu, \mbox{no
    trend}}$ is negative, then the inclusion of the trend may be
justified.  To test the significance of the reduction in $\chi^2_{\nu}$,
we employ an FAP test.  

We first employ a bootstrap method to determine
our measurement uncertainties.  We subtract the best-fit Keplerian RV curve
from the data and assume the null hypothesis --- namely that the 
residuals to this fit are properly characterized as noise and thus
approximate the underlying probability distribution function of the noise in the measurements.
We then draw from this set of residuals (with replacement) a mock set of residuals with the same
temporal spacing as the original set.  

By adding these mock residuals
to the best-fit Keplerian RV curve we produce a mock data set with the
same temporal sampling as the original data set, but with the velocity
residuals ``scrambled'' (``re-drawn'' might be a better term since
we have drawn residuals {\it with} replacement.)  It is important in
this procedure that internal errors remain associated with the scrambled residuals.
This ensures that points with error bars so
large that they contribute little to the $\chi^2_{\nu}$ sum, but
nonetheless lie far from the best-fit curve, do not gain significance
when ``scrambled'', inappropriately increasing $\chi^2_{\nu}$.

We then compare $\Delta\chi^2_{\nu}$ for our mock data set to that of
our genuine data.  By repeating this procedure 400 times, we
produce 400 mock sets of residuals and 400 values for
$\Delta\chi^2_{\nu}$.  If the linear trend is simply an artifact of the
noise, then re-drawing the residuals should not systematically improve
or worsen $\Delta\chi^2_{\nu}$.   Conversely, if a linear trend is
significant, then the null hypothesis, that the residuals to a
Keplerian are uncorrelated noise, is invalid, and re-drawing them
should worsen the quality of the Keplerian(s) plus trend fit, since
scrambling will remove evidence of the trend. 
Thus, the fraction of these sets with $\Delta\chi^2_{\nu}$ less than that
of the proper, unscrambled residuals, provides a measurement of the
false alarm probability that the residuals to a Keplerian-only fit are correlated.

\subsection{Velocity Trends and Additional Companions in Known
  Exoplanet Systems}
\label{foundtrends}
The Catalog of Nearby Exoplanets \citep{Butler06} contains 172
substellar companions  with $\msini < 24 \,\mjup$ orbiting 148 stars
within 200 pc.  Since then, at least 3 more systems have been
announced, including a triple-Neptune \citep{Lovis06}, and two
single-planet detections \citep{Johnson06}, \citep{Hatzes06}. Of these 151 systems, 24 show significant trends in
addition to the Keplerian curves of the known exoplanets.  We have
reanalyzed the radial velocities of \citet{Butler06}
to determine the significance of these trends and to find evidence for
additional trends using the FAP test described in \S~\ref{FAPtrend}.
Note that we have obtained additional data for some of these systems since
\citet{Butler06} went to press.

We confirm here 21 of the 24 trends in \citet{Butler06} to have FAPs below
1\% (2 others are in systems on which we have no data to test, and the
third is HD 11964, discussed in \S~\ref{lowamps}).  We also confirm
the trend in the 14 Her system, first announced in \citet{Naef04} and
analyzed more thoroughly in \citet{Gozdziewski06b}, and
\S~\ref{incompletes}.  We confirm the finding of \citet{Endl06} that
the trend reported in \citet{Marcy05} for HD 45350 b is not significant (FAP=0.6 and $\chi^2_{\nu}$ {\it increases} with the introduction of a trend).  

We announce here the detection of statistically significant linear
trends (FAP  $<1 \%$) around 4 stars already known to harbor a single
exoplanet:  HD 83443, GJ 436 (= HIP
57087), HD 102117, and HD 195019.  GJ 436
will be discussed more thoroughly in an upcoming work, (Maness et
al., 2006, submitted) ).  In one additional case, HD 168443, we detect a
radial velocity trend with FAP $< 1\%$ in a system already known to
have two exoplanets, indicating that a third, long-period companion
may exist.  We present the updated orbital solutions in Table~\ref{orbitupdates}. 

HD 49674 has an FAP for an additional trend of $\sim 2\%$, which is of
borderline significance when we account for the 
size of our sample:  we should expect that around 2 of our 100 systems
will prove to have FAPs $\sim 2\%$ purely by chance, and not because of an
additional companion.  We include the fit for HD 49674 with a trend in
Table~\ref{orbitupdates}, but note here the weakness of the detection.

\begin{deluxetable}{llrrcccllrcllrc}
\tabletypesize{\tiny}
\tablewidth{0pc}

\tablecolumns{15}

\tablecaption{Properties of Three Stars Hosting New Substellar Companions\label{starchar}}

\tablehead{     {HD}&{Hip
      \#}&{RA}&{Dec.}&{\bv}&{$V$}&{Distance}&{\teff}&{$\log{g}$}&{[Fe/H]}&{$v\sin{i}$}&{Mass}&{S}&{\dmv}&{jitter}\\
            & 
    & {(J2000)}     &  {(J2000)}             &             && {(pc)}
    &{(K)}       &    {(cm $\mbox{s}^{-2}$)}   &                & {$(\mps)$}
    &  {(\Msol)} & & &{(m \persec)}}
\startdata
  24040 & 17960 & 03 50 22.968   & +17 28 34.92 & 0.65 & 7.50 & 46.5(2.2) & 5853(44) & 4.361(70) & 0.206(30) & 2.39(50) & 1.18 & 0.15 & 0.65 & 5.7\\
  154345 & 83389 & 17 02 36.404   & +47 04 54.77 & 0.73 & 6.76 & 18.06(18) & 5468(44) & 4.537(70) & -0.105(30) & 1.21(50) & 0.88 & 0.18 & -0.21 & 5.7\\
  \nodata & 14810 & 03 11 14.230   & +21 05 50.49 & 0.78 & 8.52 & 52.9(4.1) & 5485(44) & 4.300(70) & 0.231(30) & 0.50(50) & 0.99 & 0.16 & 0.64 & 3.5\\
\enddata
\tablecomments{For succinctness, we express uncertainties using parenthetical notation, where the least
significant digit of the uncertainty, in parentheses, and that of the quantity
are understood to have the same place value.  Thus, ``$0.100(20)$'' indicates
``$0.100 \pm 0.020$'', ``$1.0(2.0)$'' indicates ``$1.0 \pm 2.0$'', and
``$1(20)$'' indicates ``$1 \pm 20$''.  \\ Data from columns 3--5, and 12 are from Hipparcos \citep{PerrymanESA},
columns 6--10 are from the SPOCS catalog \citep{SPOCS}, column 11 is
from \citet{Wright04}, and column 13 was derived from using the
formula in \citet{Wright05}.  Columns 6--10 for HIP 14810 were derived
with the same methods used for the SPOCS catalog.}

\end{deluxetable}

\begin{deluxetable}{l@{ }rccccccccccr}
\tabletypesize{\tiny}
\tablewidth{0pc}
\tablecolumns{13}
\tablecaption{Updated Orbital Fits for 9 Exoplanets\label{orbitupdates}}
\tablehead{ 
\multicolumn{2}{c}{Planet}  &{Per}              &{K}&{e}
&{$\omega$}          &{$T_{\mbox{p}}$}   &{trend}        
&{\msini}            &{a}                &{r.m.s.}
&{$\sqrt{\chi^2_\nu}$}&{N$_{\mbox{obs}}$}\\ 
                            &       &{(d)}&{(\mps)}&
& {($\deg$)}&{(JD-2440000)}&{(m/s/yr)}
&{($\Mjupsmall$)} &{(AU)}&{(\mps)}&
                      &                          }
\startdata 

HIP 14810 & b & 6.6742(20) & 428.3(3.0) & 0.1470(60) & 158.6(2.0) &
13694.588(40) &  & 3.91(55) & 0.0692(40) & 5.1 & 1.4 & 30  \\
HIP 14810 & c & 95.2914(20) & 37.4(3.0) & 0.4088(60) & 354.2(2.0) &
13679.575(40) &  & 0.76(12) & 0.407(23) & 3.3 & 0.77 &  21\\ 

HD 49674 & b & 4.9437(23) & 13.7(2.1) & 0.29(15) & 283 & 11882.90(86)
& 2.6(1.1) & 0.115(16) & 0.0580(33) & 4.4 & 0.62 & 39\\ 

HD 83443 & b & 2.985625(60) & 56.4(1.4) & 0.008(25) & 24 &
11211.04(82) & 2.40(79) & 0.400(34) & 0.0406(23) & 8.2 & 0.93 & 51\\ 

GJ 436 & b & 2.643859(74) & 18.35(80) & 0.145(52) & 353(24) &
11551.72(12) & 1.42(35) & 0.0682(63) & 0.0278(16) & 4.2 & 0.93 & 60\\ 

HD 102117 & b & 20.8079(55) & 11.91(77) & 0.106(70) & 283 &
10942.9(3.0) & -0.91(26) & 0.172(18) & 0.1532(88) & 3.3 & 0.83 & 45 \\ 

 HD 168443 & b & 58.11289(86) & 475.9(1.6) & 0.5286(32) & 172.87(94) &
  10047.387(34) &  & 8.02(65) & 0.300(17) & 4.1 & 0.97 & 109 \\ 
 HD 168443 & c & 1749.5(2.4)  & 298.0(1.2) & 0.2125(15) & 65.07(21) &
  10273.0(4.6)  &  & 18.1(1.5) & 2.91(17) & 4.1 & 0.97 & 109\\  

HD 195019 & b & 18.20163(40) & 272.3(1.4) & 0.0140(44)   & 222(20) &
11015.0(1.2) & 1.31(51) & 3.70(30) & 0.1388(80) & 16 & 1.5 & 154\\

\enddata
\tablecomments{For succinctness, we express uncertainties using parenthetical notation, where the least
significant digit of the uncertainty, in parentheses, and that of the quantity
are understood to have the same place value.  Thus, ``$0.100(20)$'' indicates
``$0.100 \pm 0.020$'', ``$1.0(2.0)$'' indicates ``$1.0 \pm 2.0$'', and
``$1(20)$'' indicates ``$1 \pm 20$''.}

\end{deluxetable}

\section{Constraining Long Period Companions}
\label{mapsection}
\subsection{The Problem of Incomplete Orbits}

It is difficult to properly characterize the orbit of an
exoplanet when the data do not span at least one complete revolution.
After one witnesses a complete orbit of the planet in a single-planet
system, subsequent orbits 
should have exactly the same shape (absent strong planet-planet
interactions), and so one can interpret deviations as
the effects of an additional companion.  Before witnessing one
complete orbit, one can easily misinterpret the signature of an additional
companion as it is absorbed into the orbital solution for the
primary companion.  Even when only one planet is present, small portions of
single Keplerian curves can easily mimic portions of other Keplerians
with very different orbital elements. 

\subsection{Constraining \msini\ and $P$ \label{maps}}

When an RV curve shows significant
curvature, it may be possible to constrain the minimum mass (\msini)
and orbital period of the companion.  \citet{Brown04} discussed the
problem extensively, and \citet{Wittenmyer06} studied the
significance of non-detections in the McDonald Observatory planet
search without assuming circular orbits by injecting artificial RV signals into
program data to determine the strength of a just-recoverable signal.
\citet{Wittenmyer06} reasonably assigned a broad range of eccentricities, $0 < e <
0.6$, with an upper limit they justified by the fact that over 90\% of
all known exoplanets have $e < 0.6$ \citep{Butler06}.  The presence of
this upper limit greatly limits the number of pathological solutions
to a given RV set. Below, we explore the nature of limits on mass
and period implied by a given data set and how constraining $e$ can
improve those limits. 

Since \msini, not $K$, is the astrophysically interesting quantity in
exoplanet detection, it is useful to transform into $P$, $e$, \msini\
coordinates when considering constraints.  The minimum mass of a
companion can be calculated from the {\it mass function}, $f(m)$, and
the stellar mass, according to the relation:
\begin{equation}
\label{massfn}
  f(m) = \frac{m^3 \sin^3{i}}{(m+M_*)^2} =
  \frac{P K^3(1-e^2)^{\frac{3}{2}}}{2 \pi G}
\end{equation}
where, in the minimum mass case (where $\sin{i}=1$), we set $m$ equal to
\msini.  This relation allows us to fit for the minimum mass (which
we refer to as \msini\ for brevity) eliminating the orbital
parameter $K$.

Using Eq.~\ref{massfn}, we can find the best-fit Keplerian RV curve
across $P-\msini$ space, allowing $e$, $\omega$, and $\gamma$
(the RV zero point) to vary at many fixed values of $P$ and \msini\ to map
$\chi^2$.

\section{Two New Substellar Companions with Incomplete Orbits}
\label{twonew}
\begin{figure} 
  \plotone{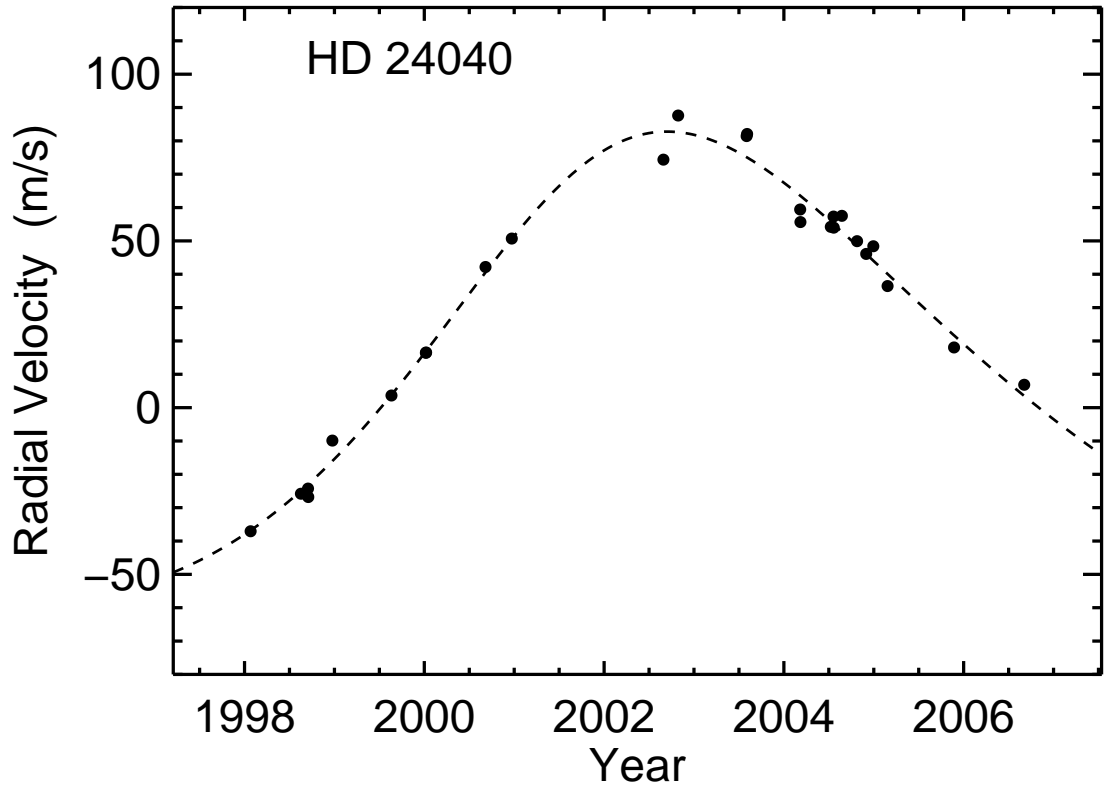}
  \caption[Radial velocity curve for HD 24040]{RV curve for HD 24040
  with data from Keck.
  The best-fit Keplerian is poorly 
  constrained due to incomplete coverage of the orbit.  The fit shown
  here is for $P =16.5$ y and $\msini = 6.9 \,\mjup$, one of a family
  of adequate solutions.\label{24040}}
\end{figure}

Here we consider HD 24040, a metal-rich ($\feh = 0.21$) G0 V star at 46 pc
(stellar characteristics summarized in Table~\ref{starchar}).  This
star shows RV variations consistent with a planetary companion with $P \sim 15$
y and $\msini \sim 7\,\mjup$ (Figure~\ref{24040}), although longer
orbital periods and minimum masses as high as $\msini \sim 30$ cannot
be ruled out.  The RV data for HD
24040 appear in Table~\ref{RVHD24040}.

Here and in \S \ref{incompletes}, we use $\chi^2$ as a merit function and infer parameters for
acceptable fits from increases of this function by 1, 4
and 9, which correspond to 1-, 2-, and 3-sigma confidence levels for
systems with Gaussian noise.  Because stellar jitter provides a source
of pseudo-random noise which may vary on a stellar rotation timescale, the
noise in RV residuals may be non-Gaussian. Thus, to the degree that
the RV residuals are non-Gauassian, the translation of
these confidence limits into precise probabilities is not
straightforward.  

\begin{deluxetable}{lrc}
\tablewidth{0pc}
\tablecolumns{3}
\tablecaption{RV Data for HD 24040\label{RVHD24040}}
\tablehead{{Time} &{Radial Velocity} & {Unc.}\\ 
{(JD-2440000)} &{(m/s)}&{(m/s)}}

\startdata
10838.773206 & -37.1 &1.4\\
11043.119653 & -25.8 &1.5\\
11072.039039 & -24.3 &1.4\\
11073.002315 & -26.8 &1.2\\
11170.876921 & -9.9 &1.4 \\
11411.092975 & 3.6  &1.6 \\
11550.824005 & 16.4 &1.3 \\
11551.863449 & 16.5 &1.2 \\
11793.136725 & 42.2 &1.2 \\
11899.945741 & 50.7 &1.1 \\
12516.065405 & 74.4 &1.3 \\
12575.951921 & 87.6 &1.5 \\
12854.115278 & 81.4 &1.2 \\
12856.115671 & 82.1 &1.1 \\
13071.741674 & 59.4 &1.2 \\
13072.799190 & 55.6 &1.3 \\
13196.130000 & 54.2 &1.3 \\
13207.120116 & 57.3 &1.3 \\
13208.125625 & 54.0 &1.2 \\
13241.091852 & 57.5 &1.2 \\
13302.947025 & 49.9 &1.3 \\
13339.945972 & 46.1 &1.2 \\
13368.865139 & 48.4 &1.2 \\
13426.809792 & 36.5 &1.1 \\
13696.904468 & 18.0 &1.1 \\
13982.061586 & 6.8  &1.2 \\

\enddata
\end{deluxetable}

\begin{figure} 
  \plotone{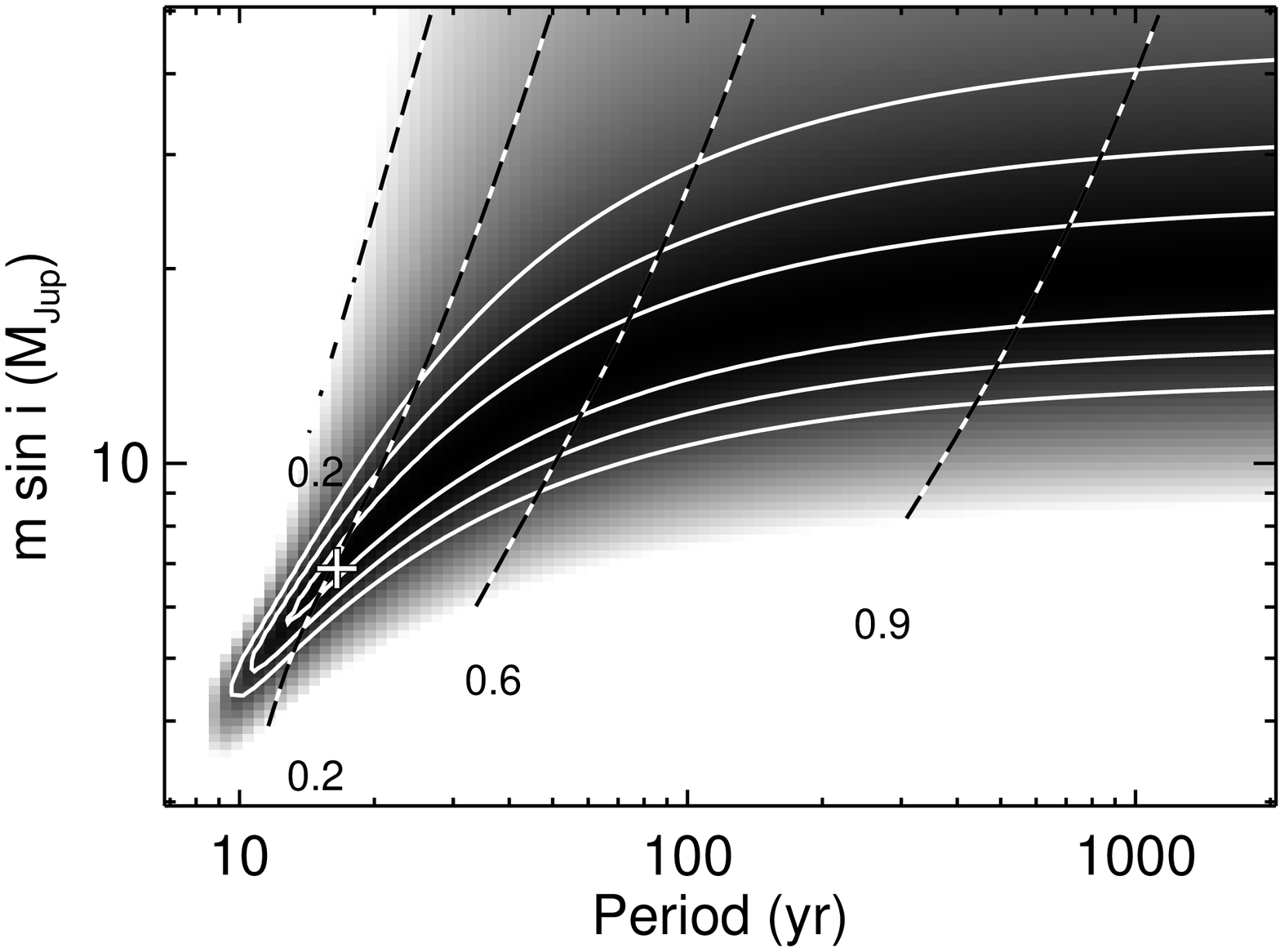}
  \caption[Contours of $\chi^2$ and $e$ of best-fit orbits
  to the radial velocity data of HD 24040]{Contours of $\chi^2$ and $e$ in $P-\msini$ space of best-fit orbits
  to the RV data of HD 24040 (Fig.~\ref{24040}), with $\chi^2$ in grayscale.  The solid contours mark
  the levels where $\chi^2$ increases by 1, 4 and 9 from the
  minimum.  The dashed contours mark levels
  of the eccentricity of 0.2, 0.6, and 0.9,
  Planets with $e > 0.6$ are rare, implying that this object is
  unlikely to have a period longer than 100 y. The orbit is largely unconstrained, but \msini\ 
  has a maximum value around 20 \mjup\ for orbits with $e < 0.6$. The position of
  the cross at 16.5 y and 6.9 \mjup\ represents the solution plotted in
  Fig.~\ref{24040}, one in a family of adequate orbital solutions.\label{24040_contour}}
\end{figure}

In this case, we have enough RV information to put an upper limit on
\msini.  Figure~\ref{24040_contour} shows $\chi^2$ for best-fit
orbits in the $P-\msini$ plan.  Fits with $P$ as low as 10 y and
\msini\ as low as 5 \mjup\ are allowed.  Interestingly, the data
(following the middle, $\chi^2 = \chi^2_{\mbox{min}} +4$ contour)
exclude orbits with $\msini > 30 \,\mjup$, providing a 
``maximum minimum-mass''.  Since without an
assumption for the eccentricity (which we will make below), we cannot exclude
orbits with \msini\ as high as 30 \mjup, there is a chance that this
companion to HD 24040 is a brown dwarf, or even stellar.

\begin{figure} 
  \plotone{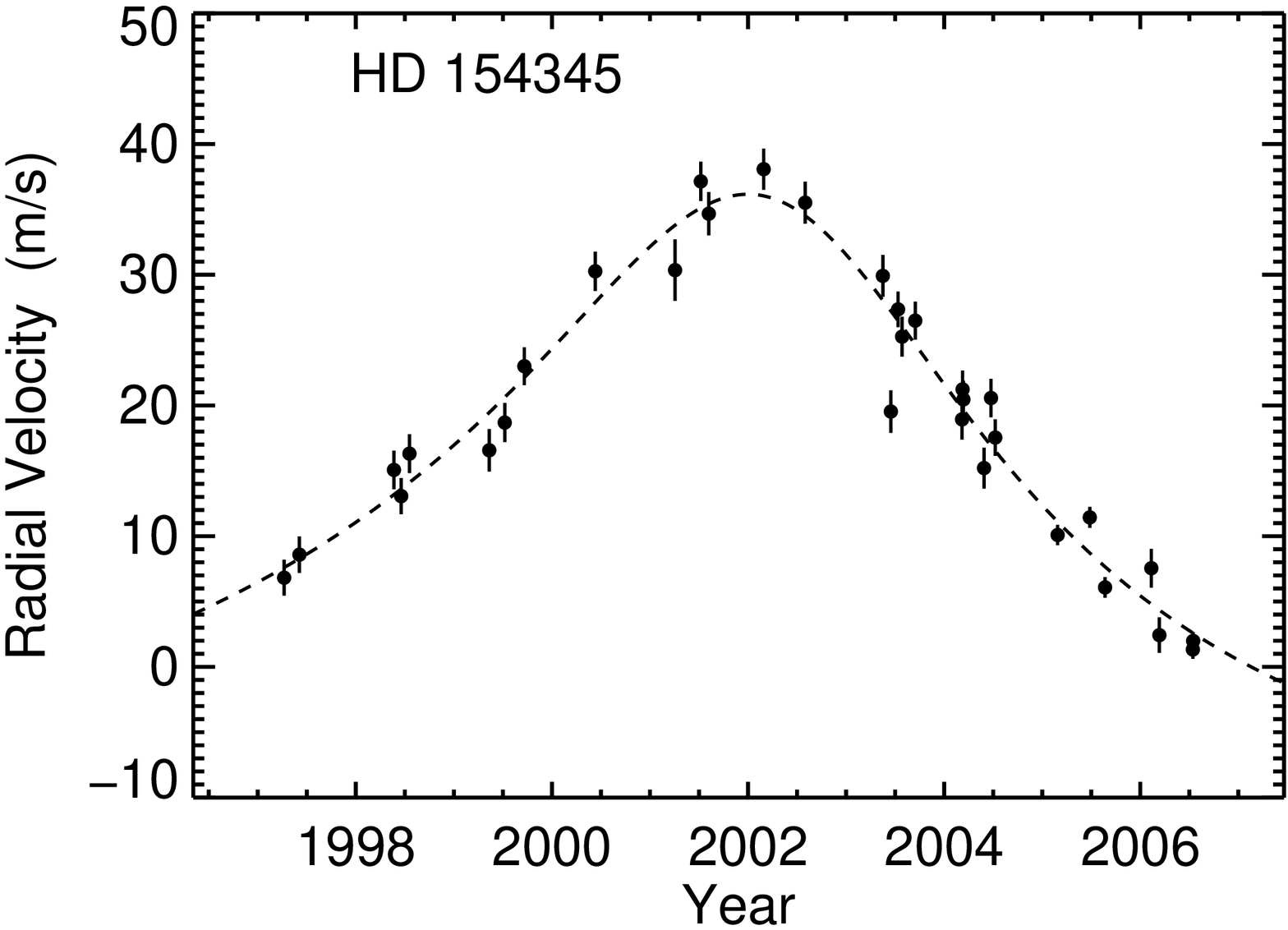}
  \caption[Radial velocity curve for HD 154345]{RV curve for HD 154345
  with data from Keck.  The best-fit Keplerian is poorly
  constrained due to incomplete coverage of the orbit.  The fit shown
  here is for $P =35.8$ y and $\msini = 2.2\, \mjup$ (marked in Fig.~\ref{154345_contour}), one in a family
  of adequate orbital solutions. \label{154345}}
\end{figure}

A similar case is HD 154345, a G8 V star at 18pc (stellar characteristics
summarized in Table~\ref{starchar}).  This star shows RV variations
remarkably similar to those of HD 24040 (Figure~\ref{154345}), but
with an amplitude about 6 times smaller.  In this case, the maximum
\msini\ is only around 10 \mjup, giving us confidence that this
object is likely a true exoplanet, and masses as low as 1 \mjup\ are
allowed.  The RV data for HD 154345 are in Table~\ref{RVHD154345}.
We summarize the orbital constraints for these objects in
Table~\ref{masslimits}.

\begin{deluxetable}{lrc}
\tablewidth{0pc} 
\tablecaption{RV Data for HD 154345\label{RVHD154345}}
\tablecolumns{3}
\tablehead{
\colhead{Time} &\colhead{Radial Velocity} & \colhead{Unc.}\\ 
\colhead{(JD-2440000)} &\colhead{(m/s)}&\colhead{(m/s)}}
\startdata
10547.110035 & 6.8 &1.4   \\
10603.955845 & 8.6 &1.4   \\
10956.015625 & 15.1 &1.5  \\
10982.963634 & 13.1 &1.4  \\
11013.868657 & 16.3 &1.5  \\
11311.065486 & 16.6 &1.6  \\
11368.789491 & 18.7 &1.5  \\
11441.713877 & 23.0 &1.4  \\
11705.917836 & 30.3 &1.5  \\
12003.078183 & 30.4 &2.3  \\
12098.916539 & 37.1 &1.5  \\
12128.797813 & 34.7 &1.7  \\
12333.173299 & 38.1 &1.6  \\
12487.860197 & 35.5 &1.6  \\
12776.985463 & 29.9 &1.6  \\
12806.951852 & 19.5 &1.6  \\
12833.801030 & 27.3 &1.4  \\
12848.772037 & 25.3 &1.5  \\
12897.776562 & 26.5 &1.5  \\
13072.046921 & 18.9 &1.6  \\
13074.077766 & 21.2 &1.4  \\
13077.128090 & 20.5 &1.4  \\
13153.943171 & 15.2 &1.6  \\
13179.992454 & 20.6 &1.5  \\
13195.819190 & 17.5 &1.4  \\
13428.162502 & 10.09 &0.78\\
13547.914433 & 11.44 &0.80\\
13604.829999 & 6.08 &0.78 \\
13777.155347 & 7.5 &1.5   \\
13807.077257 & 2.4 &1.4   \\
13931.955714 & 1.33 &0.72 \\
13932.913019 & 1.98 &0.70 \\

\enddata

\end{deluxetable}

\begin{deluxetable}{cccc}
\tablecolumns{4}
\tablewidth{0pc}
\tablecaption{Mass constraints for some substellar companions with
  incomplete orbits\label{masslimits}} 
\tablehead{{Object}  &{Per}  &{\msini} &{a} \\
   &{(y)}&{(\mjup)}&{(AU)}}
\startdata
HD 24040 b & 10 --- 100  & 5 --- 20 & 5 --- 23\\
HD 68988 c & 11 --- 60 & 11 --- 20 & 5 --- 7 \\
HD 154345 b & 7 --- 100 & 0.8 --- 10 & 4 --- 25\\
HD 187123 c & 10 --- 40 & 2 --- 5 & 5 --- 12\\
\enddata

\tablecomments{These contraints correspond to the extrema of the are
  given by $\chi^2_{\mbox{min}} +4$ contour in $P-\msini$ space for
  orbits with $e < 0.6$.}

\end{deluxetable}

\begin{figure} 
  \plotone{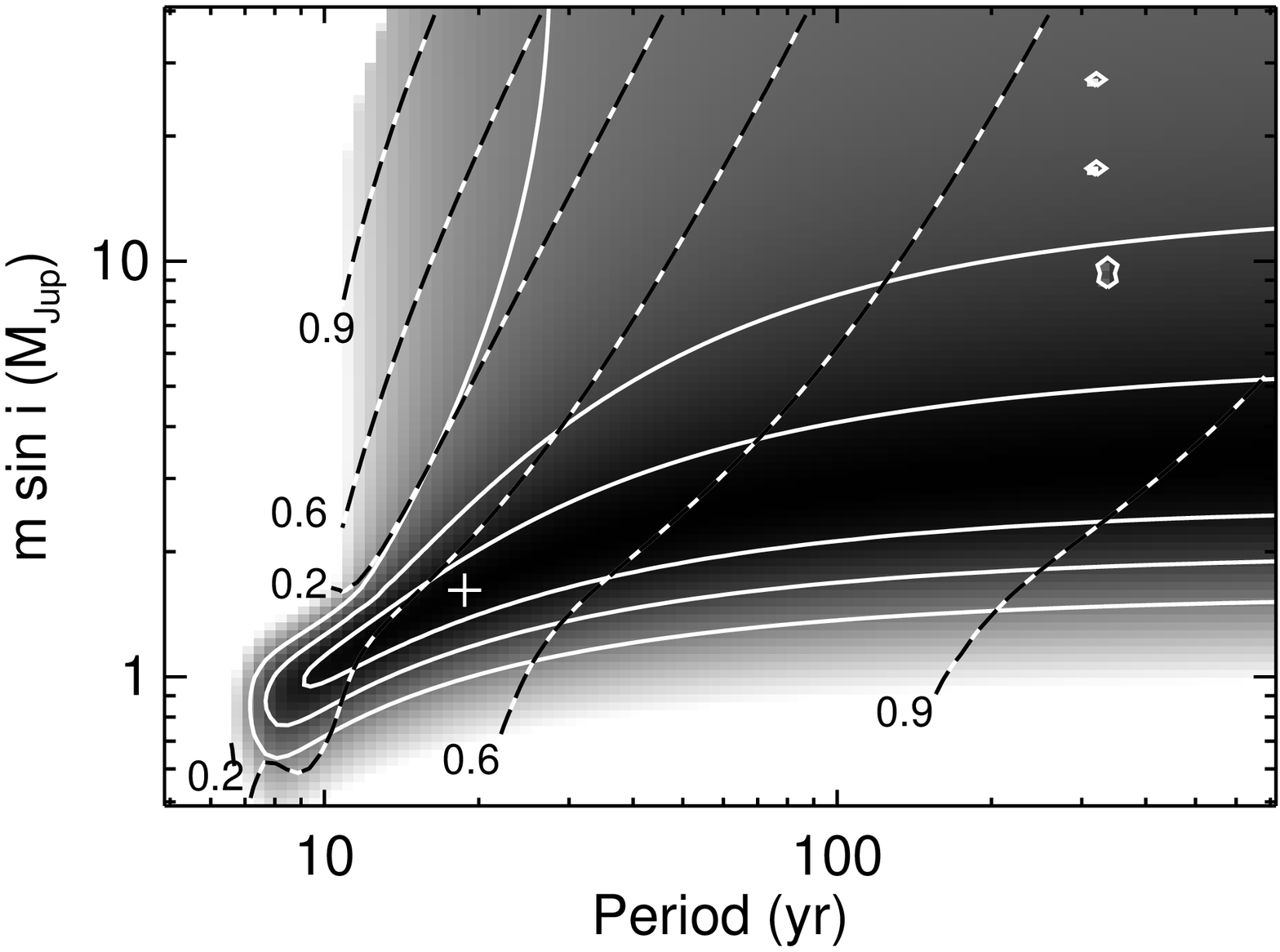}
  \caption[Contours of $\chi^2$ and $e$ of best-fit orbits
  to the radial velocity data of HD 154345]{Contours of $\chi^2$ and $e$ in $P-\msini$ space of best-fit orbits
  to the RV data of HD 154345 (Fig.~\ref{154345}), with $\chi^2$ in
  grayscale.  The solid contours mark 
  the levels where $\chi^2$ increases by 1, 4 and 9 from the
  minimum.  The dashed contours mark levels
  of the eccentricity of 0.2, 0.6, and 0.9.
  Planets with $e > 0.6$ are rare, implying that this exoplanet is
  unlikely to have a period longer than 100 y. The orbit is largely unconstrained, but \msini\ 
  has a maximum value around 10 \mjup\ for orbits with $e < 0.6$. The
  white cross at 36 y and 2.2 \mjup\ represents the solution shown in Fig.~\ref{154345}.\label{154345_contour}}
\end{figure}

We can put more stringent constraints on these orbits by noting that
ninety percent of all known exoplanets have $e < 0.6$
\citep{Butler06}.  For both HD 24040 b and HD 154345 b, the
high-period solutions all have high eccentricities (the dashed
contours in Figs.~\ref{24040_contour} and \ref{154345_contour}).  If,
following \citet{Wittenmyer06}, we therefore assume that 
$e < 0.6$ for these objects, and that the true values of \msini\ and
$P$ lie within 
the limits of the middle ($\chi^2 = \chi^2_{\mbox{min}}+4$) contour,
then we can constrain 
$5 \,\mjup < \msini < 20 \,\mjup$ and $10 < P < 100$ y for HD 24040 b, and $0.8
\,\mjup < \msini < 10 \,\mjup$ and $7 < P < 100$ y for HD 154345 b.  

\section{Mining Velocity Residuals for Additional Exoplanets}
\label{mining}
\subsection{Velocity Residuals Suggesting Additional Companions}
For exoplanetary systems in which an additional, low-amplitude
signal is not well-characterized by just a linear trend --- for instance,
where there is significant curvature (e.g. HD 13445 or HD 68988) or even multiple
orbits (e.g. GJ 876) --- a full, multi-planet fit is needed to properly
characterize the system.  In this case, we can apply an FAP analysis
similar to the one in \S~\ref{FAPtrend} testing the $(N+1)$-planet hypothesis
versus the null hypothesis of $N$ planets plus a trend plus
noise, where $N$ is the number of previously confirmed planets.
This is a much more computationally intensive procedure than that of \S~\ref{FAPtrend}, since we are
introducing 5 new, non-linear, highly covariant parameters ($P, e, \omega, T_{\mbox{p}}$,
and $K$), so we have performed only 50--100 trials.  In most cases the low-amplitude signal we seek is much weaker
than that of the known planet(s).  This means we have good initial guesses
for the orbital parameters of the established exoplanets, and that
those parameters are usually rather insensitive to those of the
additional companion, easing the difficulty of the simultaneous
11-parameter fit (16-parameter for existing double systems).

As in \S~\ref{FAPtrend}, we calculate the improvement in the
goodness-of-fit parameter,
$\Delta\chi^2_{\nu} =
\chi^2_{\nu,N+1\,\mbox{planets}}-\chi^2_{\nu,N\, \mbox{planets+trend}}$ with the introduction of an additional exoplanetary companion compared
to a fit including only an additional trend. 
We compare this reduction to that of mock data sets bootstrapped as the sum of
the best-fit solution with a trend, plus noise, drawn, with replacement,
from the residuals of the actual data to this fit.  We then construct
an FAP as the fraction of mock sets that saw a greater reduction in
$\chi^2_{\nu}$ with the introduction of an additional planet than the
genuine data set.

A low FAP for the presence of a second planet is not tantamount to the
detection of an additional exoplanet.  It is only a sign that the null
hypothesis is unlikely, i.e. that the distribution of
residuals is not representative of the actual noise in the system or
that the presumed orbital solution from which the residuals were drawn
is in error.  This would be the case if, for instance, if the
residuals are correlated due to non-Keplerian RV variations (such as
systematic errors or astrophysical jitter).

The fits discussed here are purely Keplerian and not dynamical.  In
particular, fits which produce unstable or unphysical orbits are
allowed.  More sophisticated, Newtonian fits \citep[e.g.][]{Rivera05}
would better constrain the orbits of multiple planet systems. 

\subsection{Long-Period Companions with Incomplete Orbits}
\label{incompletes}
We have identified 8 other systems in which the FAP for an additional
Keplerian vs. a simple trend is below 2\%: HD 142, HD 
13445, HD 68988, 23 Lib (= HD 134987),  14 Her,
$\tau$ Boo (= HD 120136), HD 183263, and HD 187123.  In addition, we
have identified a ninth system, HD 114783, which has a compelling
second Keplerian despite a slightly larger FAP (6\%).  We summarize the
orbital constraints for these objects in Table~\ref{masslimits}. 

--- HD 142: Most of the RV data for HD 142 show a simple linear trend
    superimposed on the known $K=34$ m/s, 350-d orbit
    \citep{Tinney02b}. HD 142 is known to have a
    stellar companion ($V=10$)
    \citep{Proveda94}, which could explain the trend.  The 
    first two data points, taken in 1998-9, are
    significantly low, producing a low FAP for curvature ($< 1\%$).
    HD 142 has \bv = 0.52, indicating it is a late F or early G star,
    suggesting it may have moderate jitter ($\sim 5$ m\persec,
    \citet{Wright05}), so we therefore view the low FAP for curvature, 
    apparently based on only two low points, with suspicion.  If the
    curvature is real, it is consistent with an exoplanet with period
    longer than the span of the observations ($P > 10$ y) with a
    minimum mass of at least 4 \mjup.

\begin{figure} 
  \plotone{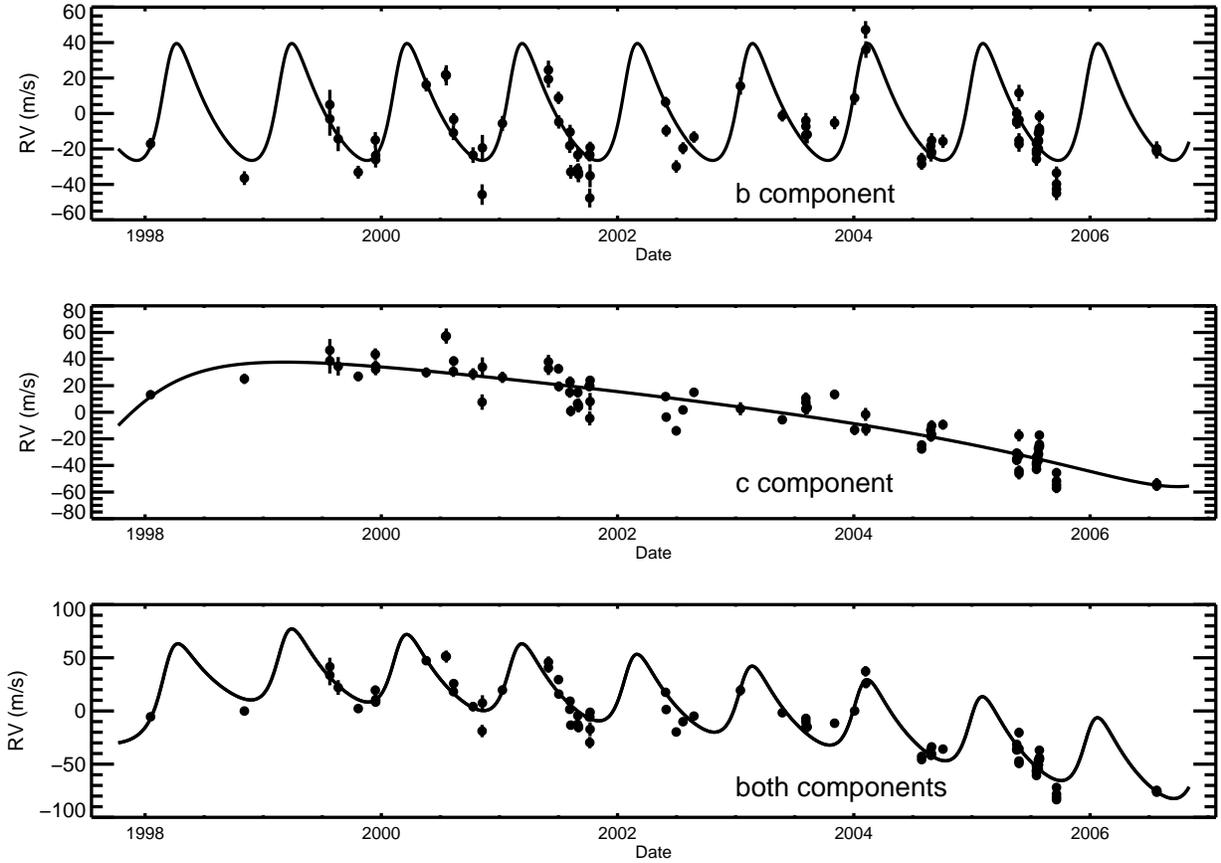}
  \caption[Radial velocity curve for HD 142]{RV curve for HD 142, a multiple companion system, with data from from AAT, .  The previously known inner
  planet has $P=350$ d, and the outer companion is poorly constrained,
  but consistent with the known stellar companion.
  The data are inconsistent with a linear trend, mostly because of the
  first two data points.
  \label{142}}
\end{figure}

--- HD 13445 (= GL 86) has a known planet with $P=15.76$ d
    \citep{Queloz00,Butler06}.Superimposed on that Keplerian velocity
    curve is a velocity trend of roughly -94 m/s/yr during the past 9
    years, apparently consistent with the brown dwarf companion
    previously reported by \citet{Els01,Chauvin06,Queloz00}.   stellar
    or brown-dwarf companion \citep{Queloz00}.  There is a a hint of
    curvature in these residuals to the inner planet, but not enough
    to put meaningful constraints on this outer object beyond that
    fact that its period is longer than the span of the
    observations ($\sim 10$ y) and $\msini > 22$ \mjup.

\begin{figure} 
  \plotone{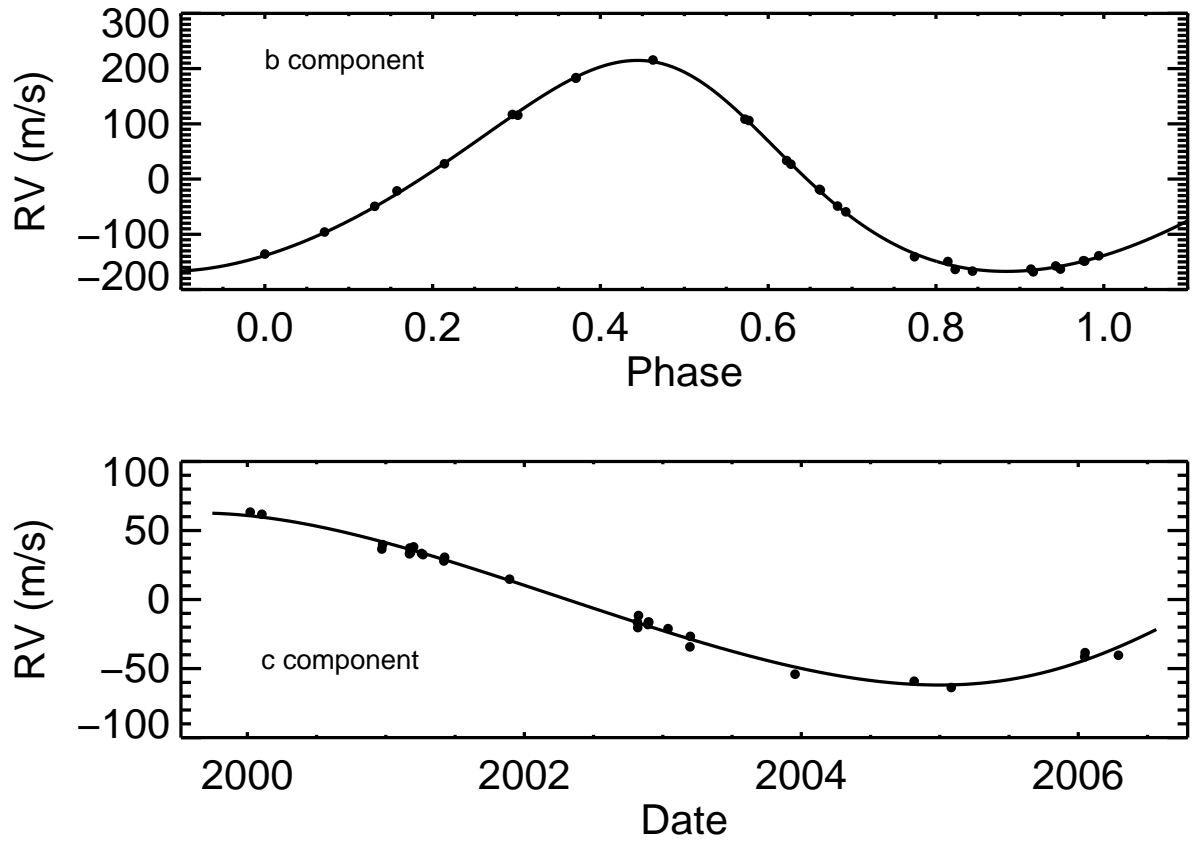}
  \caption[Radial velocity curve for HD 68988]{RV curve for HD 68988,
  a multiple-companion system, with data from Keck.  The previously known inner 
  planet has $P=6.28$ d, and the outer companion is poorly
  constrained, but likely has $\msini < 20 \mjup$ and $P < 60 y$.
  \label{68988}}
\end{figure}

\begin{figure} 
  \plotone{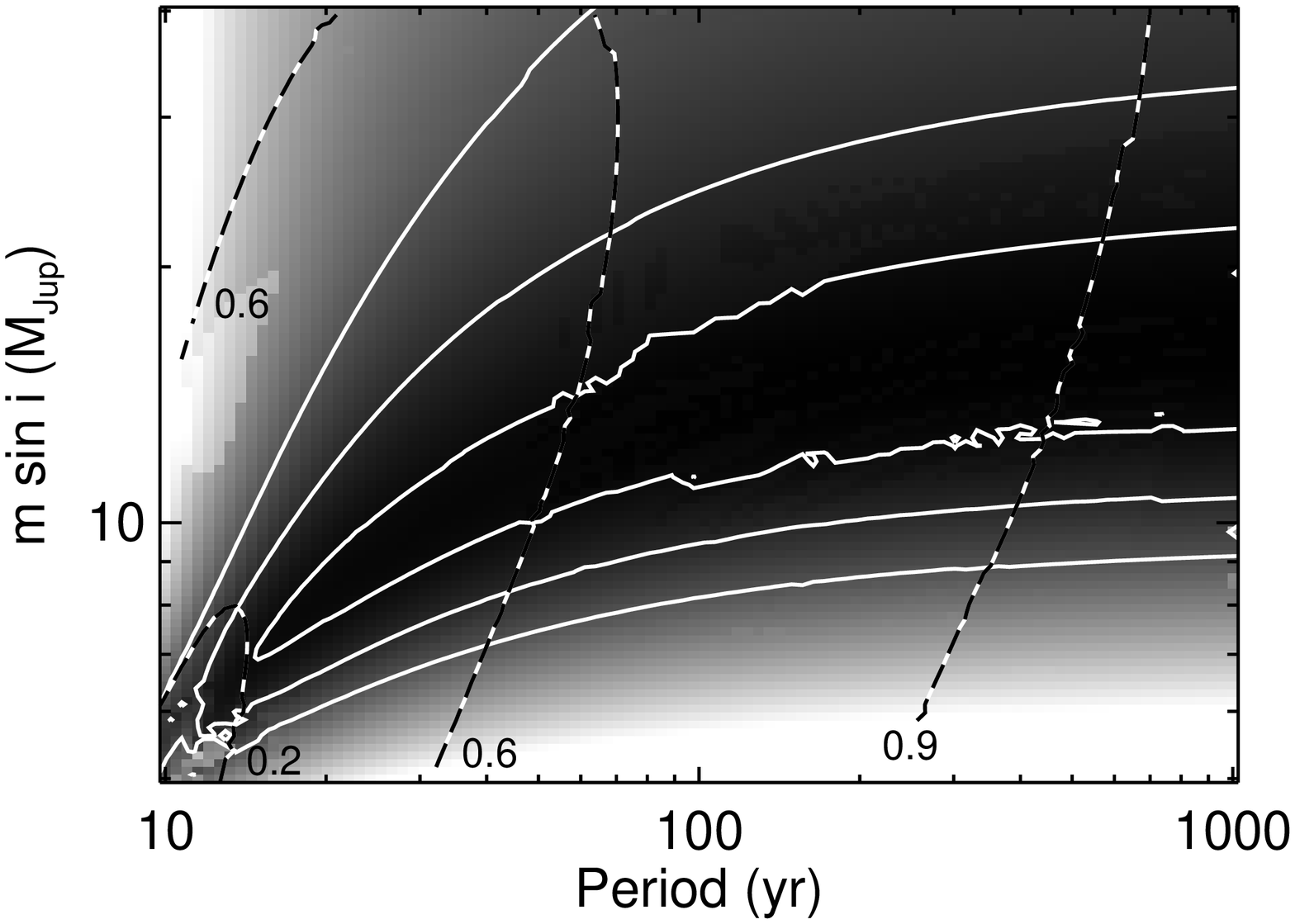}
  \caption[Contours of $\chi^2$ and $e_c$ for
  the best double-Keplerian fits to the radial velocity data of HD 68988]{Contours of $\chi^2$ and $e_c$ in $P_c-(\msini)_c$ space for
  the best double-Keplerian fits to the RV data of HD 68988
  (Fig.~\ref{68988}), with $\chi^2$ in grayscale.  The solid contours mark 
  the levels where $\chi^2$ increases by 1, 4 and 9 from the
  minimum.  The dashed contours mark levels 
  of the eccentricity of 0.2, 0.6, and 0.9.
  Assuming $e < 0.6$, we can constrain $6 \,\mjup < \msini < 20
  \,\mjup$ and $11 < P < 60$ y.   
  \label{68988_contour}} 
\end{figure}

--- HD 68988 shows definite signs of curvature in the residuals to the
    1.8\ \mjup\ inner planet (as Fig.~\ref{68988} shows).
    Fig.~\ref{68988_contour} shows the outer companion has $\msini <
    30 \mjup$, and the assumption of $e < 0.6$, using the middle
    contour, further restricts $\msini < 20 \mjup$, and $P < 60 y$. 

\begin{figure} 
  \plotone{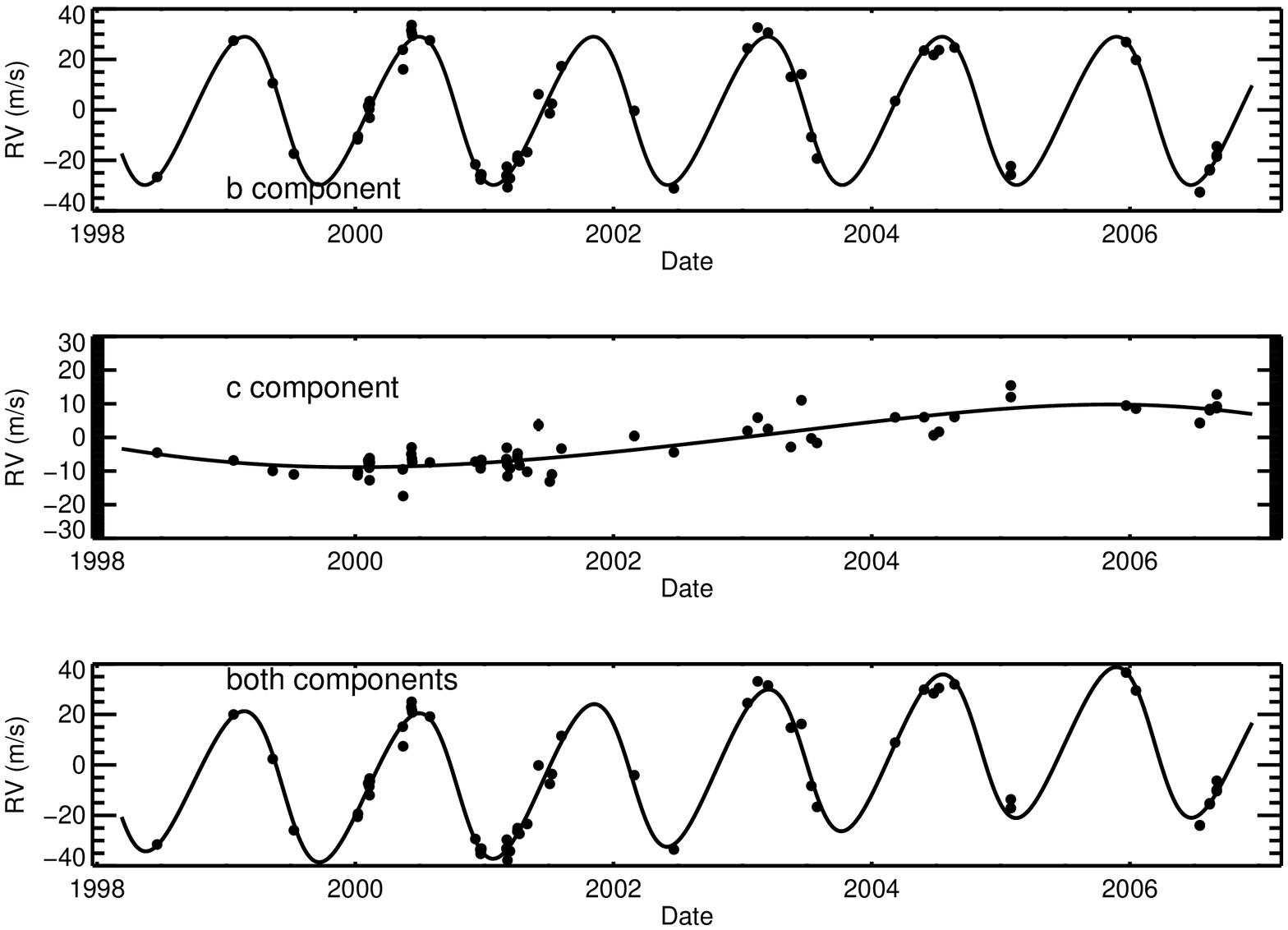}
  \caption[Radial velocity curve for HD 114783]{RV curve for HD
  114783, a multiple-companion system, with data from Keck.  The previously known inner 
  planet has $P=495$ d and $\msini = 1.1 \mjup$.  The residuals are
  only moderately inconsistent with a linear trend (FAP $= 6\%$),
  indicating that the outer companion is poorly constrained. 
  \label{114783}}
\end{figure}

--- HD 114783 shows curvature in its residuals, and may
    have experienced both an RV minimum (in 2000) and maximum (in
    2006) as the RV curve in Fig.~\ref{114783} shows.  The data are
    only moderately inconsistent with a linear trend, however (FAP =
    6\%), indicating that the outer companion's orbit is still
    underconstrained.

\begin{figure} 
  \plotone{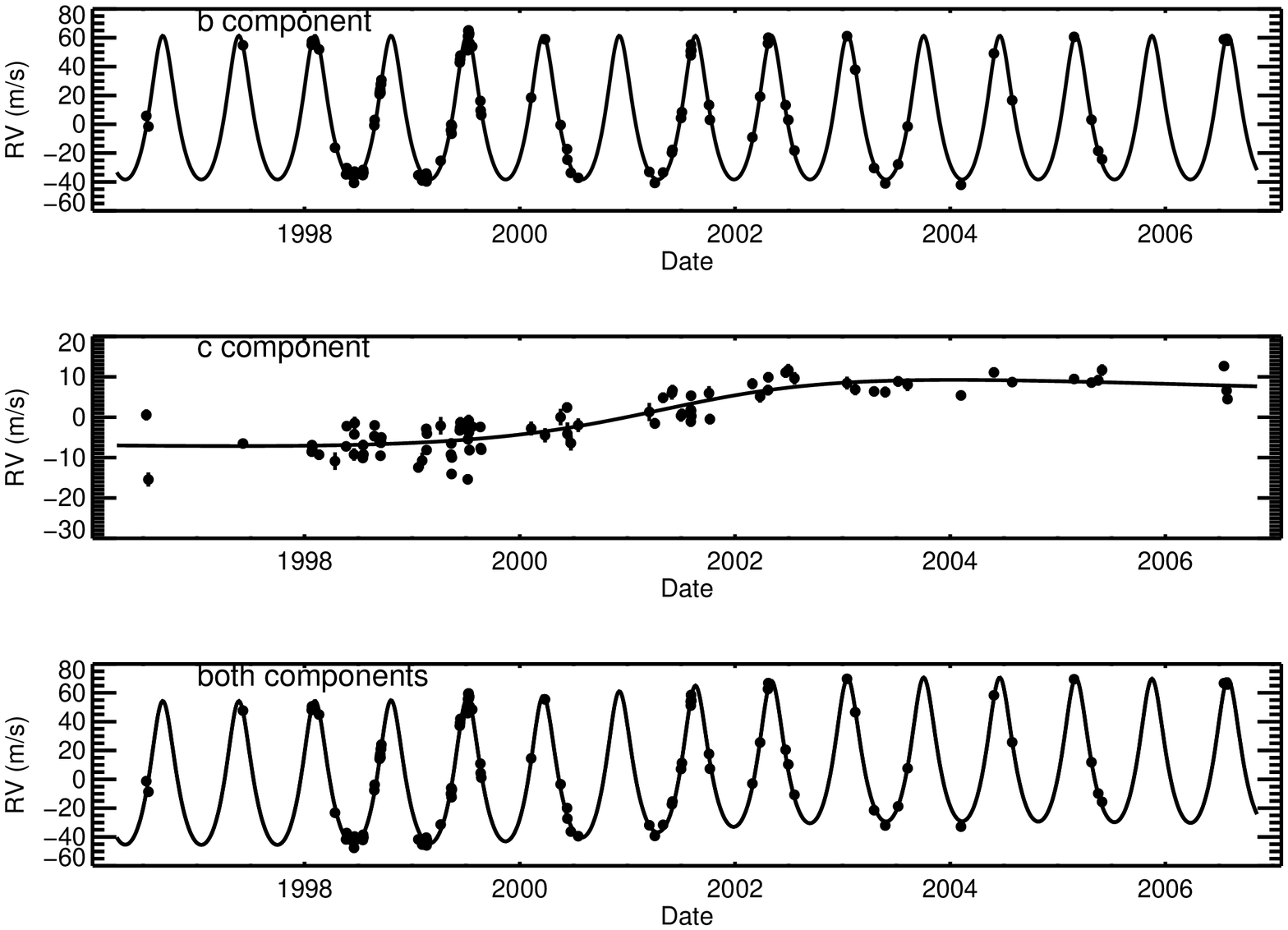}
  \caption[Radial velocity curve for 23 Lib]{RV curve for 23 Lib (= HD
  134987), a multiple-companion system, with data from Keck and AAT.  The previously known inner 
  planet has $P=258$ d and $\msini = 1.62$.  The outer companion
  is poorly constrained.  The orbital parameters of the inner planet
  are not significantly changed with a two-parameter fit.
    \label{134987}}
\end{figure}

--- 23 Lib (= HD 134987) shows signs of curvature in the residuals to
    the known inner planet .  The signal appears as a change in the
    level of otherwise flat residuals between 2000 and 2002 of 15
    m/s (see Fig.~\ref{134987}).  This suggests an outer planet on a
    rather eccentric orbit 
    which reached periastron in 2001. The small magnitude of this
    change in RV suggests a low-mass object, but the 
    incomplete nature of this orbit makes us less than certain that it
    is due to an exoplanet.  

\begin{figure} 
  \plotone{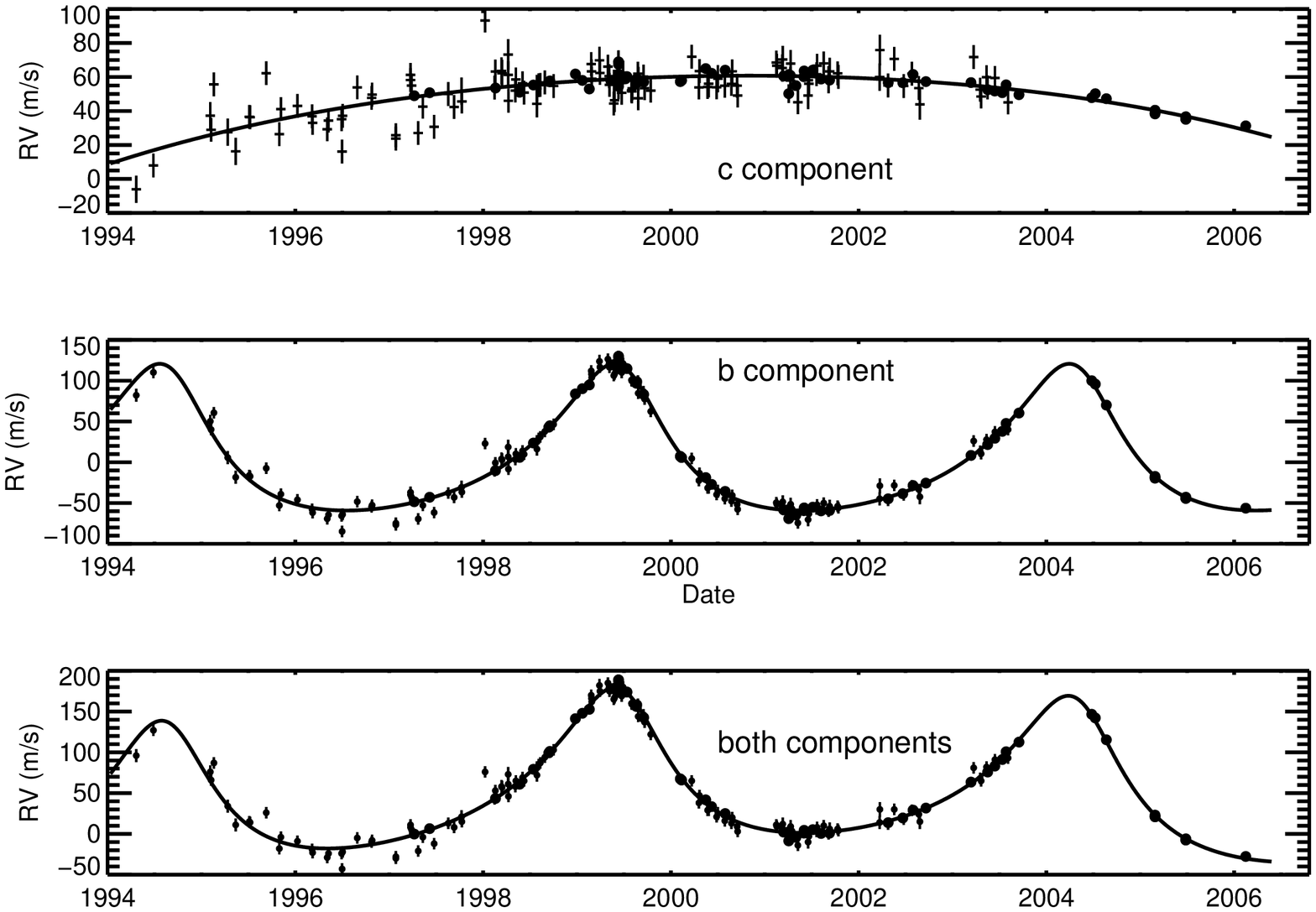}
  \caption[Radial velocity curve for 14 Her]{RV curve for 14 Her (= HD
  145675), a system with multiple 
  companions.  Crosses represent data from the ELODIE
  instrument operated by the Geneva Planet Search 
  \citep[taken from][]{Naef04}, and large filled circles represent
  data taken at Keck Observatory by the California and Carnegie Planet
  Search \citep{Butler06}.  Error bars represent quoted errors on
  individual velocities;  for some points the error bars are smaller
  than the plotted points.  The combined data set shows a long-period companion
  with $P > 12$y and $\msini > 5$ \mjup .  The previously known inner planet
  has $P=4.8$ y and $\msini = 4.9$ \mjup.  
  \label{14Her}}
\end{figure}

--- 14 Her (= HD 145645):  This star has a known trend \citep{Naef04} and has been
    analyzed by \citet{Gozdziewski06b} 
    as a possible resonant multiple system.  The previously-known
    planet has
    $\msini = 4.9$ \mjup\ and $P=4.8$ y, but the character of the second companion
    is uncertain.   Combining our data with the published ELODIE data from the
    Geneva Planet Search \citep{Naef04} (Fig.~\ref{14Her} ) provides a
    good picture of the system.  The character of the orbit of the
    outer planet is unconstrained, and several equally acceptable but
    qualitatively distinct solutions exist.  One is a long period,
    nearly circular orbit like the one shown in Fig. 10 and a mass
    near 2 \mjup.  Other solutions include a 3:1 resonance with the
    inner planet.  The next few years of observation should break this
    degeneracy.  The degeneracy may also be broken by high contrast,
    high resolution imaging, and we suggest that such attempts be made on this interesting system.

\begin{figure} 
  \plotone{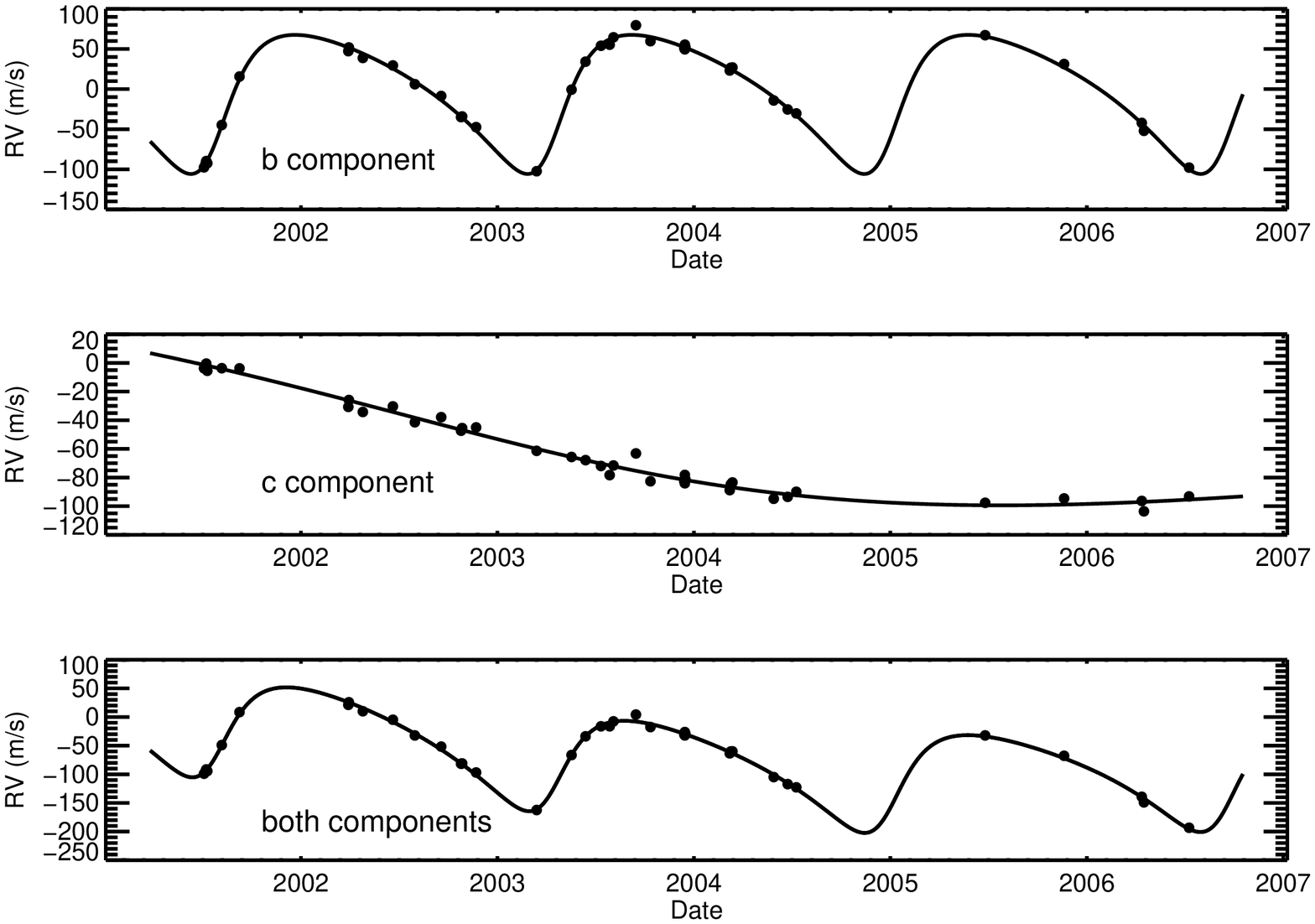}
  \caption[Radial velocity curve for HD 183263]{RV curve for HD
  183263, a multiple-companion system, with data from Keck.  The previously known inner 
  planet has $P=635$ d and $\msini = 3.8\, \mjup$, and the outer companion is poorly constrained.
  \label{183263}}
\end{figure}

--- HD 183263 shows definite signs of curvature in the residuals to
    the known inner planet (as Fig.~\ref{183263}
    shows), but too little to constrain the mass of the distant
    companion.  Fig.~\ref{183263} shows that there is little
    meaningful constraint on the orbit beyond $P > 7$ y and $\msini >
    4 \,\mjup$.  Even the assumption of $e < 0.6$ allows for $\msini >
    13$, so the planetary nature of the companion is very uncertain.

\begin{figure} 
  \plotone{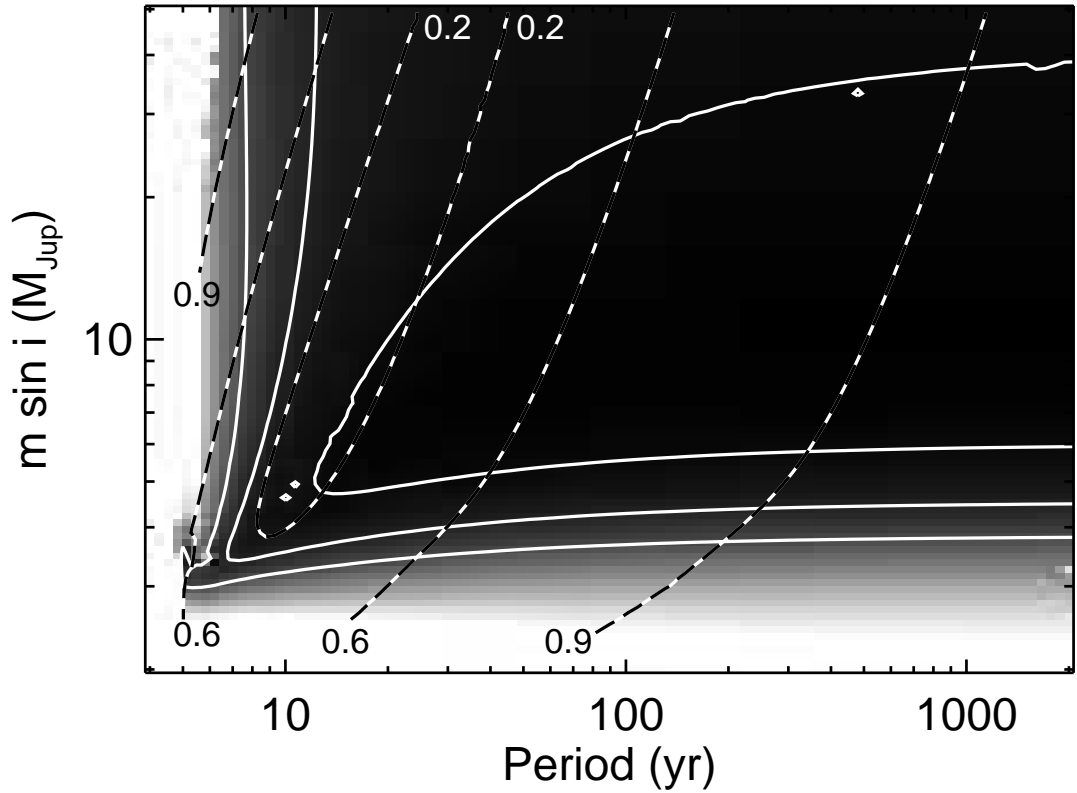}
  \caption[Contours of $\chi^2$ and $e_c$ for
  the best double-Keplerian fits to the radial velocity data of HD
  183263]{Contours of $\chi^2$ and $e_c$ in $P_c-(\msini)_c$ space for 
  the best double-Keplerian fits to the RV data of HD 183263
  (Fig.~\ref{183263}), with $\chi^2$ in grayscale.  The solid contours mark 
  the levels where $\chi^2$ increases by 1, 4 and 9 from the
  minimum.  The dashed contours mark levels   of the eccentricity of
  0.2, 0.6, and 0.9.  $P$ and 
  \msini\ for this companion are poorly constrained. 
  \label{183263_contour}} 
\end{figure}

--- $\tau$ Boo (= HD 120136) has residuals to the fit for the known
    inner planet which show evidence of a
    long-period companion which have been discussed elsewhere \citep{Butler06}.
    Analysis of the distant companion is complicated by the lower
    quality of the data during the apparent periastron in 1990.  The
    current best fit suggests a period greater than 15 years, but
    is otherwise unconstrained.  \citet{Proveda94} report that $\tau$
    Boo has a faint 
    (V=10.3) companion (sep. 5.4\arcsec) which may be the source of
    the RV residuals.

\begin{figure} 
  \plotone{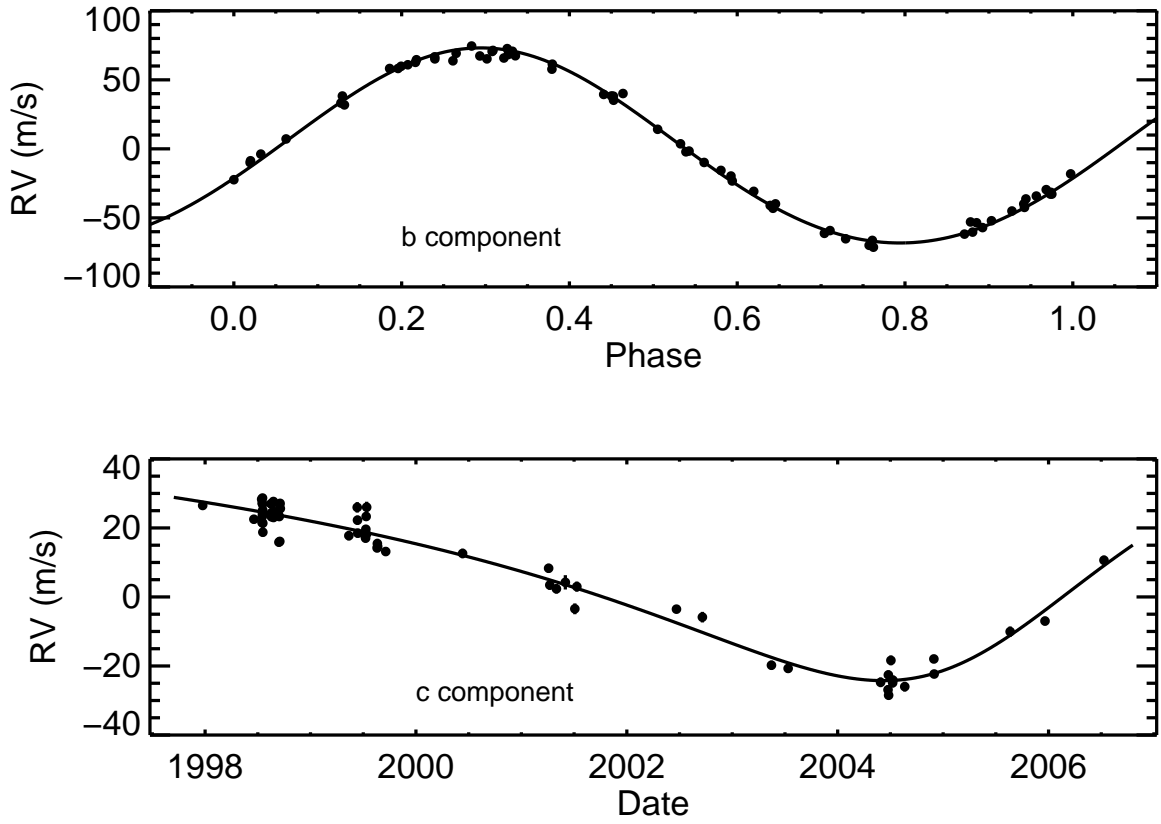}
  \caption[Radial velocity curve for HD 187123]{RV curve for HD
  187123, with data from Keck, showing the 0.5 \mjup ``Hot Jupiter'' and the outer
  companion of uncertain period and mass.\label{187123}}
\end{figure}

\begin{figure} 
  \plotone{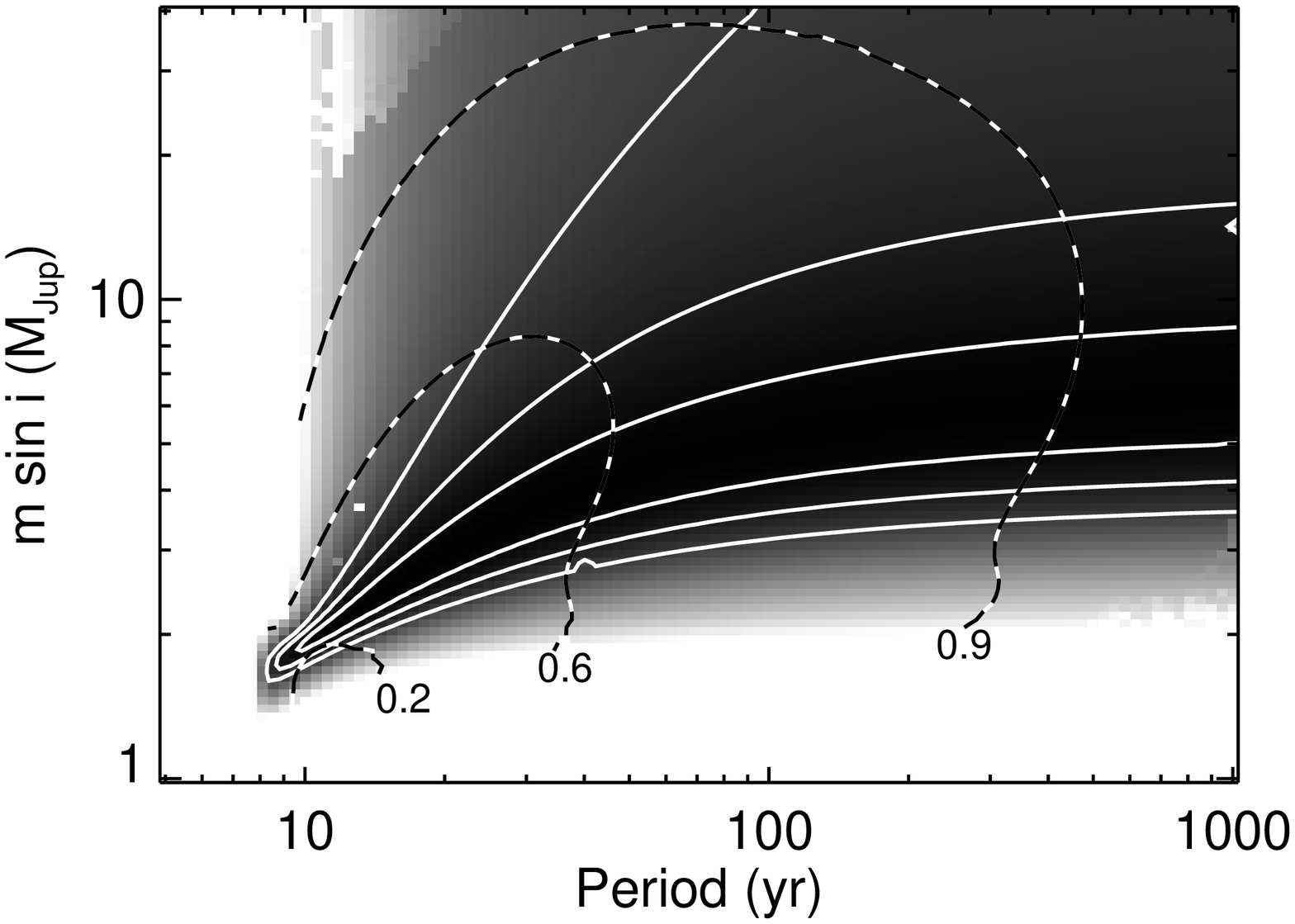}
  \caption[Contours of $\chi^2$ and $e_c$ for
  the best two-planet fits to the radial velocity data of HD 187123]{Contours of $\chi^2$ and $e_c$ in $P_c-(\msini)_c$ space for
  the best two-planet fits to the RV data of HD 187123
  (Fig.~\ref{187123}), with $\chi^2$ in grayscale.  The solid contours mark 
  the levels where $\chi^2$ increases by 1, 4 and 9 from the
  minimum.  The dashed contours mark levels 
  of the eccentricity of 0.2, 0.6, and 0.9.
  Planets with $e > 0.6$ are rare, implying that this object is
  unlikely to have a period longer than 40 y or \msini\ greater than
  5 \mjup. \label{187123_contour}}
\end{figure}

--- HD 187123 is known to host a 0.5 \mjup\ ``Hot Jupiter'' in a 3 day
    orbit \citep{Butler98}. Observations over the subsequent 8 years
    have revealed a trend of -7.3 m/s in the residuals to a one planet
    fit \citep{Butler06}. In 2001, the trend began to show signs of
    curvature, and in 2006 a it became clear that the residuals had
    passed through an RV minimum (see Fig.~\ref{187123}).
    Fig.~\ref{187123_contour} shows the $\chi^2$ and $e$ contours in
    $P-\msini$ space.  In this case, the $e=0.6$ contour and middle
    $\chi^2$ contour provide the following constraints: $2 \,\mjup <
    \msini < 5 \,\mjup$ and $10 < P < 40$ y.

\subsection{Short-Period Companions to Stars with Known Planets}
\label{lowamps}

We now consider known single-planet systems with low FAPs for second planets
whose best-fit solutions have periods shorter than the span of
observations.  We have identified eight such systems, and we discuss
them below.

Five stars appear to exhibit coherent residuals (FAP $<2 \%$)
to a one planet fit, but in all cases the best two-Keplerian fits are not
compelling (as noted in \S~\ref{foundtrends}, in a sample of 100 known
planet-bearing stars, we expect around 2 to exhibit residuals coherent
at this level purely by chance). These possible companions do not appear in
Table~\ref{orbitupdates} because the tentative nature of these signals do
not warrant publication of a full orbital solution with errors. 

--- HD 11964 was announced in \citet{Butler06} as having a planet with
    a 5.5 y orbital period and a linear
    trend.  We find an FAP for a linear trend to be 6\%, suggesting
    that while the inner planet is real, the trend is not.  We find an
    FAP for a second planet to be $< 2\%$, and a best-fit solution
    finds an inner planet with $P=37.9$d.  This star
    sits 2 magnitudes above the 
    main sequence, and the residuals to the known planet are
    consistent with the typical jitter for subgiants of 5.7 m/s
    \citep{Wright05}, so this signal could represent some sort of
    correlated noise.  This very low amplitude
    signal ($K= 5.6$ m/s) will thus require much more
    data for confirmation. 

--- HD 177830 is already known
    to have a Jupiter-mass object in a nearly circular, 1.12 y orbit.
    This remarkable system has a low FAP $< 1\%$ for a second planet vs.\ a trend.
    Two good two-planet solutions exist for this system:  the first has
    $P=111$ d and $\msini = 0.19$ \mjup, the second has $P=46.8$ and
    $\msini = 0.16$ \mjup.  This star sits more than 3.5 magnitudes above the
    main sequence, and the residuals to the known planet are
    consistent with the typical jitter for subgiants of 5.7 m/s
    \citep{Wright05}, so this signal could represent some sort of
    correlated noise. 
    
--- 70 Vir (= HD 117176) is a subgiant with a massive, 116.6 d planet on an
    eccentric orbit ($e = 0.39$).  The FAP for a second planet is 2\%,
    but the best-fit second planet is not persuasive: $P=$ 9.58 d, and
    $K= 7$ m/s.  The typical internal errors for this target
    are 5.4 m/s, making a bona fide detection of a 7 m/s planet very
    difficult.  We suspect that this signal is an artifact of stellar
    jitter, possibly due to the advanced evolution of the star. 

--- HD 164922 has a known planet with a 3.1 y orbital period.  For
    this star, the FAP for a second planet is $<1\%$.  The best 
    fit for this second planet has $P = 75.8$ d 
    and $\msini = 0.06 \mjup$.  The amplitude of this signal is
    extremely low --- only $K= 3$ m/s --- making this an intriguing but
    marginal detection. 

--- HD 210277 is already known to host a planet with a 1.2 y orbit.
    The FAP for a second planet is 2\%, and the best-fit second
    Keplerian has $K=3$ m/s signal and $P=3.14$ d, and a 2\% FAP.  The
    best-fit orbit has $e=0.5$, which is 
    unlikely given that nearly all known Hot Jupiters have $e < 0.1$
    (although the presence of the 1.2 y, $e=0.5$ outer planet could be
    responsible, in principle, for pumping an inner planet's
    eccentricity.)  The extremely low amplitude of this 
    planet makes the exoplanetary nature of this signal very
    uncertain.

Three additional stars with low FAPs are of a very early
spectral type (F7--8): HD 89744 \citep{Korzennik00}, HD 108147 \citep{Pepe02}, and HD
208487 \citep{Tinney05}.  Their low activity
yields a low jitter in the estimation of \citet{Wright05}, but this is
likely underestimated due to poor statistics:  the California and
Carnegie Planet Search has very few stars of this spectral type from
which to estimate the jitter.   For HD 89744 and HD 108147 we suspect
that, the low 
FAP of $< 2\%$ is an artifact of coherent noise, since in our judgment neither
case shows a compelling evidence of a second Keplerian of any period. 

HD 208487 has a very low FAP ($< 1\%$) despite the modest r.m.s of the
residuals to a one planet fit of 8 m/s.  We suspect that stellar
jitter is the likely source of these variations.  This star was
discussed by \citet{Gregory05}, who applied a 
Bayesian analysis to the published RV data, concluding that a
second planet was likely, having $P = 998^{+57}_{-62}$ d and
$\msini \sim 0.5 \mjup$.  \citet{Gozdziewski06} also studied the
published data, and suggested a planet with $P=14.5$ d. We note
here two plausible solutions apparent in our data.
The first, with $P \sim 1000$ d and $\msini \sim 0.5$, is
consistent with the solution of \citet{Gregory05}.  We also find,
however, an additional solution of equal quality with $P=28.6$ d
(double the period of \citet{Gozdziewski06})
and $\msini = 0.14 \,\mjup$.  This second solution has a period
uncomfortably close to that of the lunar cycle (we often see
this period in the window function of our observations 
due to our tendency to observe during bright time).  For both
solutions $K=10$ m/s.  We reiterate that we feel that the early
spectral type of this star alone can account for the observed RV residuals.

\citet{Gozdziewski06} analyzed our published RV data and found a low
FAP for the existence of a second 
planet in orbit around HD 188015.  Their FAP, however, is measured against
a null hypothesis of a single Keplerian plus noise, thus ignoring the
linear trend.  We find that $\Delta\chi^2_{\nu}$, the improvement of
the goodness-of-fit parameter with the introduction of a second
Keplerian versus a trend to be very small --- in fact 60\% of
our mock data sets showed greater improvement.  We therefore find no motivation
to hypothesize the existence of an additional, short-period
planet; the single planet and trend announced in \citet{Marcy05} are
sufficient to explain the data.

\citet{Gozdziewski06} also found a low FAP for a second planet in
orbit about HD 114729, with a period of 13.8 d.  Using the data set
from \citet{Butler06}, which contains 3 recent RV measurements taken
since the publication of \citet{Butler03} (their 
source of RV data), we find no such signal, and a large FAP for a
second planet.  We suspect our results may differ
because the additional data provide for a slightly better fit to the
known exoplanet, changing the character of the residuals and
destroying the coherence of the spurious 13.8 d signal.
\section{HD 150706}

HD 150706 b, a purported 1.0\,\mjup\ eccentric planet at 0.8 AU, was
announced by the Geneva Extrasolar Planet Search Team 
(2002, Washington conference ``Scientific Frontiers in Research
in Extrasolar Planets"; Udry, Mayor \& Queloz, \citeyear{Udry03b}) and
appears in \citet{Butler06}; however, there is no refereed discovery
paper giving details. 

We have made eight precision velocity measurements at Keck observatory
from 2002 through 2006.  These velocities show an RMS scatter of 12.1
m/s, inconsistent with the reported 33m/s semi-amplitude of HD 150706 b.
The RMS to a linear fit is 8 m/s, which is adequately explained
by the expected jitter for a young (700$\pm$300 Myr) and active star
like HD~150706.  We therefore doubt the existence of a 1.0\,\mjup
eccentric planet orbiting HD 150706 at 0.8 AU.

\section{Discussion}
As noted in \S~\ref{foundtrends}, prior to this work 24 of the 150 nearby
stars known to host exoplanets (including 14 Her and excluding HD
150706) show significant trends in 
their residuals and 19 host well-characterized multiple planet
systems.  One of these trends is likely spurious (HD 11964), and at
least 3 others may be due to stellar or brown dwarf companions
(HD 142, HD 13445, and $\tau$ Boo).   We have announced here the detection of
an additional 5 trends for known planet-bearing stars, 2 new single systems,
and one new multiple system (HIP 14810, which appears as a single-planet
system in \citet{Butler06}).  We have also confirmed that the
previously announced trends for HD 68988 and HD 187123 are likely due
to planetary-mass objects.  This brings the total number of
stars with RV trends possibly due to planets to 22, the number of
known multiple-planet systems to 22, and the number of nearby
planet-bearing stars to 152.  This means 
that 30\% of known exoplanet systems show significant evidence of multiplicity.
Considering that the mass distribution of planets increases steeply
toward lower masses \citep{Marcy_Japan_05}, our incompleteness must be
considerable between 1.0 and  0.1 Jupiter-masses.  Thus, the actual
occurrence of multiple planets among stars having one known planet
must be considerably greater than 30\%. 

From an anthropocentric perspective, this frequency of multiplicity
suggests that in some respects, the Solar System is not such an
aberration.  Our Sun has 4 giant planets, and it appears that such
multiplicity is not uncommon,although circular orbits are.

From a planet-hunting perspective this result is
quite welcome as well, since it means that the immediate future of RV
planet searches looks bright.  As our temporal baseline
expands, we will become sensitive to longer-period planets. Our search
is just becoming sensitive to true Jupiter analogs with 12 year orbits
and 12 m/s amplitudes.  A true Saturn analog would require 15 more
years of observation.  As our precision improves we will become
sensitive to lower-mass planets, which may be the richest domain for planets
yet. 

\acknowledgements

The authors would like to thank Kathryn Peek for obtaining the crucial
16 Apr 2006 RV measurement of HIP 14810, and Simon O'Toole
and Alan Penny for their assistance.  The authors also thank the
anonymous referee for a thorough and constructive report.

This research is based on observations obtained
at the W. M. Keck Observatory, which is operated jointly by the
University of California and the California Institute of Technology.
The Keck Observatory was made possible by the generous financial
support of the W. M. Keck Foundation. The authors wish to recognize
and acknowledge the 
very significant cultural role and reverence that the summit of Mauna
Kea has always had within the indigenous Hawaiian community.  We are
most fortunate to have the opportunity to conduct observations from
this mountain.

This research has made use of the SIMBAD database, operated at CDS,
Strasbourg, France, and of NASA's Astrophysics Data System
Bibliographic Services, and is made possible by the generous support
of Sun Microsystems, NASA, and the NSF.

\end{document}